\documentclass[11pt]{article}
\usepackage[margin=1in]{geometry}  %

\usepackage{graphicx}
\usepackage{upgreek}
\usepackage{multicol,multirow}
\usepackage{amsmath,amssymb,amsfonts}
\usepackage{amsthm}
\usepackage[figuresright]{rotating}
\usepackage{appendix}
\usepackage[authoryear]{natbib}
\usepackage{ifpdf}
\usepackage[T1]{fontenc}
\usepackage{newtxtext}
\usepackage{newtxmath}
\usepackage{textcomp}
\usepackage{xcolor}
\usepackage{float}
\usepackage{siunitx}
\usepackage[colorlinks,allcolors=blue]{hyperref}
\usepackage[most]{tcolorbox}
\usepackage{amsmath}
\usepackage{newtxmath}

\newtcolorbox{boxtext}{
  colback=gray!10,    %
  colframe=black,     %
  boxrule=0.5pt,      %
  arc=2mm,            %
  left=6pt, right=6pt, top=6pt, bottom=6pt,
}

\newcommand{\appref}[1]{Appendix~\ref{#1}}

\renewcommand{\abstract}[1]{%
  \begin{origabstract}#1\end{origabstract}%
  \vspace{1em}%
  \noindent\textbf{Keywords:} Natural ventilation, Airflow networks, Large eddy simulation, Urban canopy flow, Building energy modeling, Discharge coefficient\par
  \vspace{1em}}

\definecolor{jourcolor}{cmyk}{1,0.57,0.01,0.38}
\hypersetup{
    colorlinks,
    citecolor=jourcolor,
    filecolor=jourcolor,
    linkcolor=jourcolor,
    urlcolor=jourcolor
}

\theoremstyle{definition}

\title{Improving Airflow Network Models of Natural Ventilation: Geometry-Specific Pressures and Fluctuation Corrections}

\author{
Nicholas Bachand\textsuperscript{1}, 
Hesam Salehipour\textsuperscript{2}, 
Catherine Gorlé\textsuperscript{1} \\
\\
\textsuperscript{1}Civil and Environmental Engineering Department, Stanford University, Stanford, USA \\
\textsuperscript{2}Autodesk Research, Toronto, Canada
}

\date{}

\begin{document}

\maketitle

\maketitle

\abstract{
Building energy models typically predict natural ventilation rates using airflow networks, which solve a system of orifice equations with exterior pressures as boundary conditions. This study investigates two improvements to these models: (i) geometry-specific exterior pressures obtained from large eddy simulations (LESs) and (ii) analytical corrections accounting for the nonlinear effects of pressure fluctuations on mean ventilation rates. We evaluate these improvements against coupled interior-exterior LESs of flows through 5824 windows, 2688 skylights, and 2752 interiors across idealized urban canopies. Replacing standard ASHRAE pressure coefficients with pressures from exterior-only LESs reduces the RMSE of predicted ventilation rates by approximately 50\%, largely by capturing urban canopy flow effects that are poorly represented by standard shelter factors. Independently of the pressure source, mean-pressure airflow network predictions exhibit a nonlinear relationship with the LES ventilation rates. To address this discrepancy, we derive fluctuation corrections for mean-dominated and fluctuation-dominated flow regimes. These corrections eliminate the systematic nonlinearity and further reduce RMSE by 10--40\% for most opening types and ventilation configurations. For windows, the combined improvements from geometry-specific surface pressures and fluctuation corrections increase the coefficient of determination $R^2$ from approximately 0.4 to 0.9. The results demonstrate the potential to accurately predict natural ventilation rates across different interior layouts while accounting for urban canopy effects and pressure fluctuations, using exterior-only LES pressure information.}

\section{Introduction}
\label{sec::Intro}

Natural ventilation relies on wind and buoyant forces to passively drive outside air through buildings. By bringing in cooler outside air, natural ventilation can reduce or eliminate the need for mechanical cooling during hotter times of day. However, the potential for natural ventilation can be difficult to predict. Interior flow fields experience an array of complex processes including flow through multiple openings, through various interior layouts, and with different turbulent characteristics. Additionally, the interior flow is directly coupled to the exterior flow field, which has its own features and turbulent properties~\citep{gorle_urban_2026}.

Building energy models (BEMs)  often represent natural ventilation with 1-D pressure-driven airflow networks derived from potential flow through sharp openings~\citep{ramponi_energy_2014}. These models require time-averaged pressures specified at openings to predict time-averaged ventilation rates~\citep{etheridge_natural_2012}. The resulting predictions can be inaccurate for two reasons: 1) the specified pressures do not reflect the true surrounding wind field and 2) the true ventilation rate is fluctuating. 

These airflow networks often assume pressures corresponding to flow around simplified geometries. The American Society of Heating, Refrigerating and Air-Conditioning Engineers (ASHRAE) reports surface pressures from wind tunnel tests on isolated model-scale buildings~\citep{americansocietyofheatingrefrigeratingandair-conditioningengineers_ashrae_2025}. Using isolated buildings has been common practice, as there is no obvious choice for a surrounding canopy that would reflect the diversity of urban development~\citep{hussain_wind_1980}. Instead, models commonly consider canopy effects by adjusting surface pressures measured on isolated buildings~\citep{costola_overview_2009, grosso_wind_1992, chiesa_pythonbased_2019}. In reality, the urban canopy changes the local flow field in complex ways which bulk adjustments do not reflect~\citep{bachand_simulating_2026, vanhooff_effect_2010, king_investigating_2017, adachi_numerical_2020, vargiemezis_predicting_2026, gorle_urban_2026}. The ASHRAE guidelines point out these limitations and recommend using building-specific pressures for complex building shapes, significant nearby obstacles~\citep{americansocietyofheatingrefrigeratingandair-conditioningengineers_ashrae_2025}, or later design stages~\citep{simmonds_ashrae_2021}. However, without more rigorous analysis, the ASHRAE pressures remain a common option for BEMs. The engineering reference for EnergyPlus, a very widely used BEM, explicitly points to the ASHRAE surface pressures in its airflow network section~\citep{doe_energy_2025}. If not user-specified, the built-in calculation uses area-averaged surface pressures~\citep{swami_correlations_1988}, which are even less precise than the 2-D surface distributions in ASHRAE.
                                
The second source of error arises because traditional airflow networks neglect turbulent fluctuation effects when estimating mean ventilation rates. %
Fluctuations induce errors in time-averaged quantities because the ventilation rate scales as the square-root of the pressure difference.  This square-root error has largely gone unaddressed. \citet{etheridge_building_1996} report no systematic investigations of square-root errors for building ventilation, but mention similar efforts in the field of pressure-based flow meters for industrial applications~\citep{p.m.downing_effect_1975}. 

In fact, the literature on pressure-based flow meters has investigated fluctuation induced errors since the early 20th century~\citep{hodgson_commercial_1917, judd_effect_1922, georgek.burgess_circular_1926}. In 1921, J.L. Hodgson introduced a quantity, later termed the Hodgson number, to describe the square-root error~\citep{packard_discussion_1922, hodgson_laws_1929, zanker_review_1970}. Assuming quasi-steady flow without temporal-inertia or compressibility, the square-root error for a differential pressure measurement of pulsating flow would be constant at a constant Hodgson number~\citep{hodgson_orifice_1925}. \citet{mottram_behaviour_1971}
later showed that, for incompressible flow with temporal-inertia effects, the square-root error depends only on the ratio of the standard deviation of velocity fluctuations over the mean velocity magnitude. If temporal-inertia effects are then excluded, the square-root error can be expressed using the ratio of the standard deviation of pressure fluctuations over the mean pressure differential. The review paper by \citet{gajan_influence_1992} also discussed these square-root error formulations alongside many experimental results. These same expressions are included in the 2018 ISO standard on measuring pulsating flows~\citep{iso_measurement_2018}, and are only valid when the flow's standard deviation is significantly smaller than its mean.

It is worthwhile mentioning that the relationship between fluctuations and mean ventilation rates is different from the well-studied relationship between fluctuations and net air exchange, which includes airflow both in and out of an opening~\citep{etheridge_unsteady_2000a, haghighat_modelling_2000, allard_airflow_1992, etheridge_natural_2012}. Airflow networks predict mean ventilation rates and are therefore not well suited for situations with important bi-directional flow. Even with bi-directional flow, mean rates may sometimes be more appropriate than net rates for describing ventilation. Transient airflow reversals do not necessarily mix with the interior air~\citep{cockroft_ventilation_1976}, contributing less to heat transfer or pollutant removal. 

Our objective is to investigate two complementary methods for improving airflow network predictions of natural ventilation, each addressing one of the aforementioned two drawbacks. The first method uses geometry-specific surface pressures as airflow network boundary conditions. These pressures derive from large-eddy simulations (LESs) of exterior flow fields. We compare the geometry-specific airflow network predictions with those using ASHRAE~\citep{americansocietyofheatingrefrigeratingandair-conditioningengineers_ashrae_2025} pressures, quantifying performance against measured ventilation rates from LESs of coupled exterior-interior airflow. The second method corrects ventilation rates to account for fluctuation errors. For this, we derive analytical correction terms associated with two ventilation regimes: a mean-dominated and a fluctuation-dominated regime. In the mean-dominated regime, we arrive at the same correction introduced by \citet{mottram_behaviour_1971} for flow meters. For the fluctuation-dominated regime however, we propose a new correction term. We present a smooth piecewise model that spans these two regimes continuously. This model characterizes the two regimes using a non-dimensional fluctuation intensity $I$, a ratio comparing the standard deviation of the fluctuations to the mean. We verify these fluctuation corrections through Monte Carlo simulations and validate them against the coupled interior-exterior LESs.

The remainder of this work is organized as follows. We split up methods into two sections. Section~\ref{sec::VentAndAirflow} introduces the LESs used for validation and the standard airflow network model. Section~\ref{sec::AFNImprovements} then describes our proposed methods to improve airflow networks. These improvements are geometry-specific surface pressures and corrections for fluctuation errors. Section~\ref{sec::Results} compares the results of 8 airflow network formulations and quantifies the impacts of our proposed improvements. Section~\ref{sec::Discussion} discusses our recommended airflow network formulations and Section~\ref{sec::Conclusion} summarizes the conclusions.

\section{LES Validation Data Set and Standard Airflow Network Model}
\label{sec::VentAndAirflow}

In this section we introduce the methodological foundations which we will later build upon for our airflow network improvements. Section~\ref{ssec::M-ValidationData} discusses the LESs used for validation while Section~\ref{ssec::M-StandardAFN} covers the standard airflow network models against which we evaluate our proposed model improvements. We discuss both the formulation of standard airflow networks (Section~\ref{sssec::M-StandardAFNForm}) and typical pressure boundary conditions (Section~\ref{sssec::M-ASHRAE}).

\subsection{LES Validation Data}
\label{ssec::M-ValidationData}

\begin{figure}[htb]
    \centering
    \includegraphics[width=0.99\textwidth]{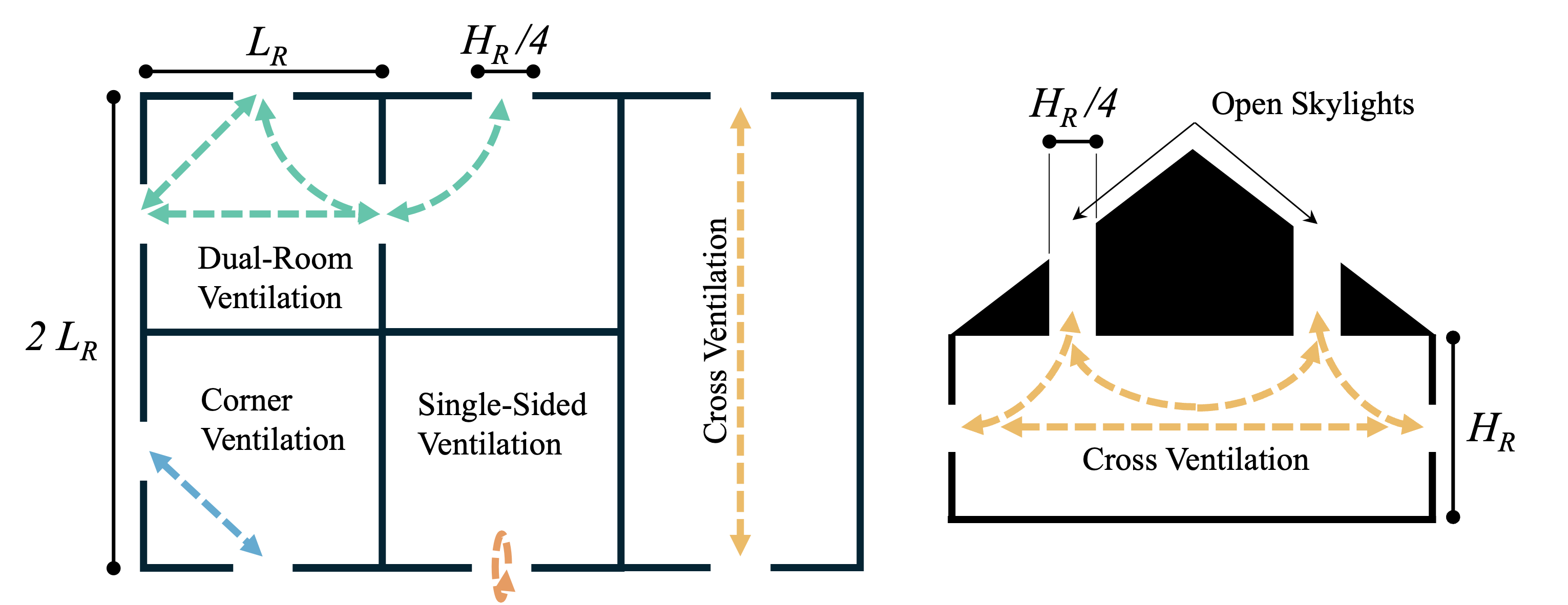}
    \caption{Building interiors and possible ventilation pathways.} 
    \label{fig:ventLayout}
\end{figure}

All LESs run with the low Mach formulation of Cadence's CharLES solver~\citep{cascadetechnologies_charles_2022}. In this work we build on LESs initially described in \citet{bachand_simulating_2026}, which details the governing equations, numerics, domain setup, computational grid, time step, boundary conditions, atmospheric boundary layer (ABL), and ventilation measurement methods. In the present work, we performed simulations using the same methods unless otherwise noted. 

\begin{figure}[htb]
    \centering
    \includegraphics[width=0.9\textwidth]{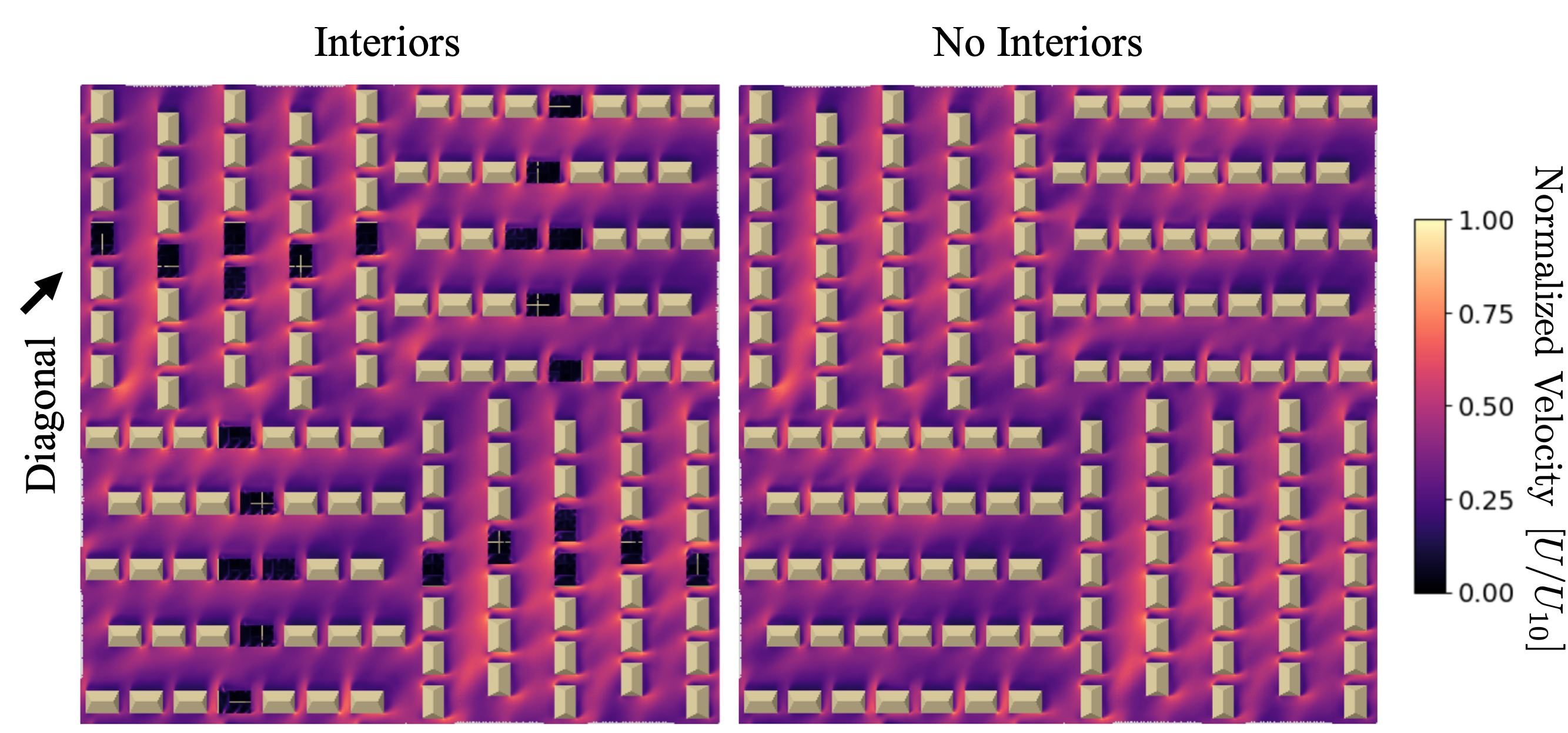}
    \caption{Domain-scale flow through the high density canopy for a coupled interior-exterior LES (left) and exterior-only LES (right). Each LES contained four identical quadrants rotated by $90^\circ$. With this configuration, each LES captured the quadrant geometry under four wind angles~\citep{bachand_simulating_2026}.} 
    \label{fig:flowFields}
\end{figure}

\paragraph{Flow Rates from Coupled Interior-Exterior LESs}
The coupled interior-exterior LESs in \citet{bachand_simulating_2026} placed houses with the same geometry in a variety of flow conditions by perturbing wind speed (2 and 4 m/s), wind direction (every 45°), indoor-outdoor temperature difference (0 and 5 $^\circ$C), canopy density (high and low), and relative location within the canopy. Each quadrant (Figure~\ref{fig:flowFields}) contained six simulated interiors located throughout: five identical houses without skylights and one house with skylights. The house interiors contain four separate sections with different ventilation characteristics: cross, corner, dual-room, and single-sided ventilation. The plan view in Figure~\ref{fig:ventLayout} shows these four ventilation types and corresponding window locations while sketching possible flow paths through the building interior. Each simulated condition was initialized twice for a total of at least 18 flow-through periods of averaging time. We average the measured and modeled ventilation across these two (very similar) runs.

This work includes an additional set of simulations configured identically to those in \citet{bachand_simulating_2026} but with every house containing skylights. Each square room has a skylight centered in the ceiling. The rectangular cross-ventilated room has two skylights centered over each half, as shown by the elevation view in Figure~\ref{fig:ventLayout}, for a total of six skylights per house. These simulations were only run once (as the different initializations in the earlier LESs gave very similar results) for a minimum of nine flow-through periods of averaging time.

As \citet{bachand_simulating_2026} details, each building interior initializes with an indoor-outdoor temperature difference applied to both the indoor air volume and surfaces. The outdoor air volume and surfaces initialize to a zero reference temperature. All surfaces keep their temperature during the simulation, but the indoor air volume equilibrates during spin-up, balancing surface temperature flux and air exchange. Using the temperature in each cell, we calculate the buoyant force using a $3.4 \times 10^{-3} K^{-1}$ thermal expansion coefficient and 10 $\text{m/s}^2$ gravitational acceleration.

From these LESs we measure the mean ventilation rate through openings and interiors. For openings, we take both the spatial average across the window and temporal average across simulation time. Therefore, any spatial or temporal flow reversals reduce the mean ventilation rate. This mean quantity is different from the net quantity used in \citet{bachand_simulating_2026}, which counted flow reversals as additional ventilation. For interiors, we measure the total ventilation rates into and out of the interior through all connected openings. The flows in and out of each interior are theoretically equal by continuity, and we account for small numerical differences by using the average of the two as the interior flow rate. We exclude single-sided rooms without skylights where the mean ventilation rate is always zero.

\subsection{Standard Airflow Network Model}
\label{ssec::M-StandardAFN}
Airflow network models predict wind- and buoyancy-driven flows by solving a system of 1-D equations. This section describes our baseline pressure boundary conditions, the equations for flow through sharp openings, and the steadiness assumptions that underpin airflow network models.

\subsubsection{ASHRAE Pressures}
\label{sssec::M-ASHRAE}
For our baseline pressure boundary condition, we use pressure coefficients ($C_p$) for a low-rise building with a rectangular floor plan and pitched roof, as provided in the ASHRAE Fundamentals Handbook~\citep{americansocietyofheatingrefrigeratingandair-conditioningengineers_ashrae_2025} and the ASHRAE Design Guide for Natural Ventilation~\citep{simmonds_ashrae_2021}. These coefficients are based on the work of \citet{holmes_wind_1994} and correspond to a building geometry similar to ours.  We account for surrounding buildings using ASHRAE shelter classes and the corresponding shelter factors from \citet{walker_field_1998}, which scale the reference wind speed and thus the predicted ventilation rates. 

Table~\ref{tab:ashrae_pressures} lists the pressure coefficients for eight wind directions. Three representative windows are shown; the remaining windows are rotational transformations of these configurations. We note that ASHRAE defines its pressure coefficients relative to the wind speed at room height $H_R$, whereas our simulations use the wind speed at 10 m, $U_{10}$. We convert the ASHRAE coefficients to this reference using a logarithmic atmospheric boundary layer profile with an aerodynamic roughness length of 0.3 m~\citep{stull_introduction_2003}. We only apply ASHRAE pressures to windows, as the available roof pressure coefficients are insufficiently detailed for predicting skylight ventilation. 
\begin{table}[htbp]
\centering
\caption{Pressure coefficients for low-rise buildings from the ASHRAE fundamentals handbook~\citep{americansocietyofheatingrefrigeratingandair-conditioningengineers_ashrae_2025}. The first eight columns show the pressure coefficients for different wind angles, while the last four columns show the prescribed sheltering effects. HDC and HDF are the high density canopy shelter class and shelter factor, respectively, while LDC and LDF are the low density canopy shelter class and shelter factor. Pressure coefficients are relative to the wind speed at room height $H_R$.}
\resizebox{\textwidth}{!}{%
\begin{tabular}{lrrrrrrrrrrrr}
\hline
Window case & \multicolumn{8}{c}{$C_p$ for Wind Angle} & \multicolumn{4}{c}{Sheltering Effects} \\
\cline{2-9} \cline{10-13}
 & $0^\circ$ & $45^\circ$ & $90^\circ$ & $135^\circ$ & $180^\circ$ & $225^\circ$ & $270^\circ$ & $315^\circ$ & HDC & HDF & LDC & LDF \\
\hline
Corner Window Along Width   & 0.70 & 0.20 & -0.30 & -0.30 & -0.14 & -0.30 & -0.50 & 0.45 & 5 & 0.30 & 4 & 0.50 \\
Corner Window Along Length   & -0.40 & 0.50 & 0.75 & 0.20 & -0.20 & -0.30 & -0.43 & -0.55 & 4 & 0.50 & 3 & 0.70 \\
Single-Sided Window  & -0.20 & 0.35 & 0.80 & 0.35 & -0.20 & -0.40 & -0.43 & -0.40 & 4 & 0.50 & 3 & 0.70 \\
\hline
\end{tabular}%
}
\label{tab:ashrae_pressures}
\end{table}

Table~\ref{tab:ashrae_pressures} also lists the shelter classes and factors for the high density (HD) and low density (LD) canopies.
We apply a single shelter factor to all windows in each airflow network calculation, selected based on the window with the largest $C_p$. Following the ASHRAE classification, shelter class 3 represents buildings across the street, class 4 represents closer obstacles more than one building height away, and class 5 represents immediately adjacent obstacles. In the low density canopy we assign classes 3 and 4 across streets and across side yards, respectively. In the high density canopy we adjust these classes up to 4 and 5.

\subsubsection{Model Formulation}
\label{sssec::M-StandardAFNForm}

Airflow networks account for the driving forces provided by wind and buoyancy using surface pressures and indoor-outdoor temperature differences. We use indoor-outdoor temperature differences measured from the coupled interior-exterior LESs. To measure these temperatures, we take horizontal averages across each interior. We calculate the associated buoyant contribution using the same thermal expansion coefficient ($3.4 \times 10^{-3} K^{-1}$) and gravitational acceleration (10 $\text{m/s}^2$) as in the LESs. Because all windows are the same height, buoyancy only affects networks with skylights.

Given pressure and temperature boundary conditions, airflow networks solve a set of 1-D flow equations derived from the potential flow theory. These equations are all versions of the orifice equation, which gives the ventilation rate ($q_i$) through a sharp opening as:

\begin{equation}
q_i = S_i C_{d,i} A_i \sqrt{\frac{2 |\Delta p_i|}{\rho}}
\label{eq:qSO}
\end{equation}
where $S_i$ is the sign of the pressure difference across the window, $C_{d,i}$ is the discharge coefficient, $A_i$ is the window area, $\Delta p_i$ is the pressure difference across the window, and $\rho$ is the density of air~\citep{etheridge_natural_2012}. $\Delta p_i$ can be decomposed into multiple components:
\begin{equation}
\Delta p_i = \Delta p_{0} - \Delta \rho_i \cdot g z_i + p_{w,i}
\end{equation}
where $\Delta p_{0}$ is the hydrostatic pressure difference, $\Delta \rho_i$ is the buoyancy-induced density difference across an opening, $g$ is acceleration due to gravity, $z_i$ is the opening height, and  $p_{w,i}$ is the wind-driven pressure. The hydrostatic pressure difference $\Delta p_0$ is calculated to satisfy conservation of mass. For a room with $n$ openings:
\begin{equation}
\sum_{i=1}^{n} q_{i}(\Delta p_0) = 0
\label{eq:p0argmin}
\end{equation}
Numerically, $\Delta p_0$ is found by minimizing the magnitude of the left-hand side of~\eqref{eq:p0argmin}.
Assuming potential flow, and that the streamlines exiting a sharp opening eventually become parallel, $C_d$ is analytically derived to be 0.611~\citep{etheridge_natural_2012}.

\subsubsection{Mean-Pressure and Steadiness Assumptions}
\label{sssec::M-Assumptions}
While the orifice equation does not include dynamic effects, it is often applied in scenarios where there is fluctuating flow. Airflow networks commonly use a mean-pressure form of the orifice equations~\eqref{eq:qSO}, directly substituting the time-averaged pressure to calculate the time-averaged ventilation rate:

\begin{equation}
q_{\text{MP}} = S_\text{MP}C_{d} A \sqrt{\frac{2 |\overline{\Delta p}|}{\rho}}
\label{eq:qAFN}
\end{equation}
$q_{\text{MP}}$ is the predicted time-averaged ventilation rate, $\overline{\Delta p}$ is the time-averaged pressure, and $S_\text{MP} = \operatorname{sgn}(\overline{\Delta p})$. 

\section{Airflow Network Model Improvements}
\label{sec::AFNImprovements}
This section introduces two methodological improvements to airflow networks: geometry-specific surface pressure boundary conditions (Section~\ref{ssec::M-GeoPressures}) and corrections for mean-pressure fluctuation errors (Section~\ref{ssec::M-Corrections}). We verify these corrections through Monte Carlo analysis (Section~\ref{ssec::M-MonteCarlo}), describe their implementation within airflow networks (Section~\ref{ssec::M-CorrectedAFN}), and explore various options for formulating the airflow network setups (Section~\ref{ssec::M-Evaluated}).

\subsection{Geometry-Specific Surface Pressures}
\label{ssec::M-GeoPressures}

To obtain geometry-specific surface pressures, we performed a second set of LESs with only the exterior flow, representing all houses as closed solid volumes. Figure~\ref{fig:flowFields} shows the domain-scale flow through the high density canopy for both an original coupled interior-exterior LES and new exterior-only LES. The exterior-only LESs use the same mesh refinement described in \citet{bachand_simulating_2026} and include all wind conditions from the coupled interior-exterior LESs.  We do not modulate indoor-outdoor temperature differences since the interior is not simulated. These simulations restart from the flow states of the corresponding interior-exterior LESs, providing a matched exterior-only analogue for each coupled run.%

\begin{figure}[htb]
    \centering
    \includegraphics[width=0.9\textwidth]{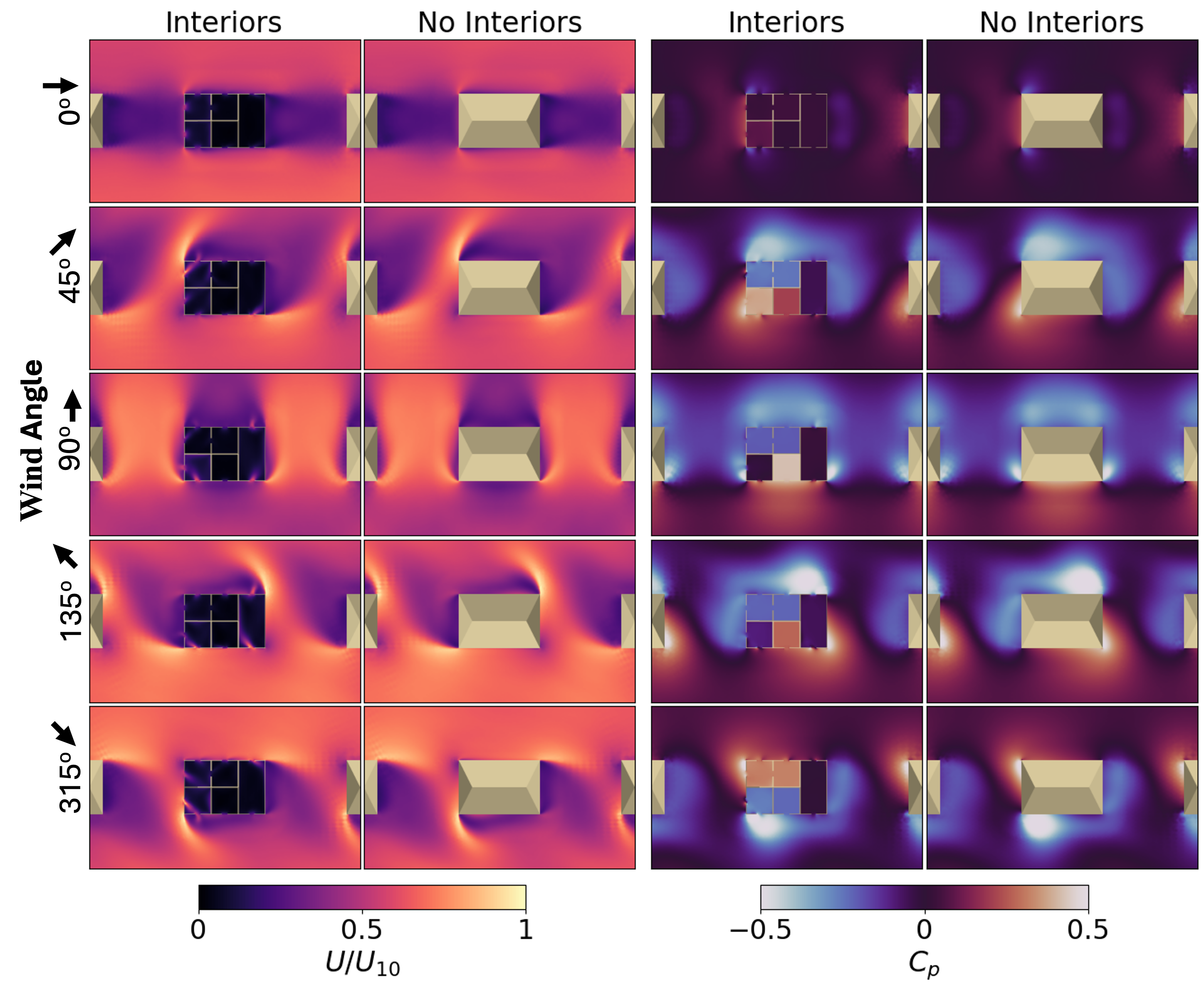}
    \caption{Mean velocity magnitude and pressure coefficients at window-center height for five wind angles. Alternating columns show coupled interior-exterior LESs and exterior-only LESs. All visualizations show the quadrant-center house in the low-density canopy.} 
    \label{fig:flowAndPressure}
\end{figure}

From these exterior-only simulations, we extract the surface pressures at the locations corresponding to the windows and skylights in the coupled interior-exterior simulations. Figure~\ref{fig:flowAndPressure} compares the coupled interior-exterior and exterior-only LESs. The left two columns show the mean velocity field while the right two columns show the mean pressure field. Noticeable differences are confined to the direct vicinity of windows. Away from the openings, the velocity and pressure fields are nearly identical, confirming that the ventilation flow does not significantly alter the surrounding exterior flow field. This supports using surface pressures from the exterior-only simulations as the external pressure inputs to the airflow network models. For each opening, we spatially average the time-averaged surface pressure over the corresponding opening area to obtain an opening-specific pressure coefficient, $C_p$.

\subsection{Fluctuation Corrections}
\label{ssec::M-Corrections}

An alternative to the mean-pressure assumption outlined in Section~\ref{sssec::M-Assumptions} is to treat the orifice equation as describing the flow instantaneously. This quasi-steady assumption holds if dynamic effects, namely inertia and compressibility, are small at the timescale of the fluctuation. \citet{etheridge_unsteady_2000} shows that the quasi-steady assumption often holds for natural ventilation, particularly for short openings. \citet{etheridge_unsteady_2000a} further demonstrates that dynamic effects have little impact on time-averaged flow rates. Even though turbulence in the urban canopy contains a wide range of scales, large-scale fluctuations have the biggest impact on ventilation rates and occur at timescales where dynamic effects tend to be small. 

One way to express the quasi-steady form is through Reynolds decomposition: $q = \overline{q} + q'$ and $\Delta p = \overline{\Delta p} + \Delta p'$, where $\overline{(\cdot)}$ denotes the time-averaged mean while $(\cdot)'$ denotes the deviations from this mean:
\begin{equation}
\overline{q} + q' = S C_{d} A \sqrt{\frac{2 |\overline{\Delta p} + \Delta p'|}{\rho}}
\label{eq:inst_flow}
\end{equation}
$S = \operatorname{sgn}(\overline{\Delta p} + \Delta p')$ is the sign function that ensures the flow rate $q$ has the same sign as the instantaneous pressure drop $\Delta p$. Exactly applying the quasi-steady assumption to an arbitrary pressure time series requires the full time series. This requirement makes the quasi-steady assumption largely unused in practice~\citep{etheridge_natural_2012}. However, in the remainder of this section we show that, with certain assumptions, we can describe the quasi-steady solution using only statistical moments of the pressure or ventilation time series.

The fluctuation corrections are different depending on the relative size of the fluctuations compared to the mean. 
Section~\ref{sssec::M-F-MeanDominated} presents the correction for fluctuations that are small compared to the mean, while Section~\ref{sssec::M-F-FluctDominated} presents the correction for fluctuations that are large compared to the mean. \appref{app::MeanDominated} and~\appref{app::FluctDominated} give the full derivations. In Section~\ref{sssec::M-F-Transition} we discuss the transition between regimes, and in Section~\ref{sssec::M-F-Piecewise} we combine the regimes into a single non-dimensional piecewise model spanning all fluctuation levels.
\subsubsection{Mean-Dominated Regime}
\label{sssec::M-F-MeanDominated}

When fluctuations are small relative to the mean, we assume the pressure difference never changes sign, i.e.\ $\operatorname{sgn}(\overline{\Delta p} + \Delta p') = \operatorname{sgn}(\overline{\Delta p})$. Squaring and time-averaging~\eqref{eq:inst_flow} under this assumption gives: 
\begin{equation}
(\overline{q})^2 + \overline{(q')^2} = (q_{\text{MP}})^2
\label{eq:final_avg_flow_sqrd}
\end{equation}
where $q_\text{MP}$ is the mean-pressure ventilation rate traditionally produced by airflow networks. \appref{app::MeanDominated} gives the full derivation. Rearranging~\eqref{eq:final_avg_flow_sqrd} for $\overline{q}$ gives:
\begin{equation}
\overline{q} = q_{\text{MP}}\sqrt{1 - I^2}
\label{eq:qAFN_qstd}
\end{equation}
where $I$ is the fluctuation intensity through the opening:
\begin{equation}
I(q)=\frac{\sqrt{\overline{(q')^2}}}{|q_{\text{MP}}|} = \frac{\sigma_q}{|q_{\text{MP}}|}
\label{eq:FQ_qstd}
\end{equation}
We can alternatively approximate $I$ as a function of pressure fluctuations. A second-order binomial expansion of~\eqref{eq:inst_flow} about $|\Delta p'| \ll |\overline{\Delta p}|$ recovers~\eqref{eq:qAFN_qstd} with $I$ expressed in pressure terms, valid for small $I$ (\appref{app::MeanDominated}):
\begin{equation}
I(\Delta p) \approx \frac{\sqrt{\overline{(\Delta p')^2}}}{2|\overline{\Delta p}|} = \frac{\sigma_{\Delta p}}{2|\overline{\Delta p}|}
\label{eq:FQ_pstd}
\end{equation}

\subsubsection{Fluctuation-Dominated Regime}
\label{sssec::M-F-FluctDominated}

If fluctuations are much larger than the mean we can no longer assume that the sign of pressure does not reverse. Applying a first-order binomial expansion about small $|\overline{\Delta p}/\Delta p'|$, the leading term time-averages to zero assuming a symmetric distribution of $\Delta p'$, leaving $\overline{q}$ as (\appref{app::FluctDominated}):
\begin{equation}
\overline{q}
\approx
\frac{q_{\text{MP}}}{2H}
\label{eq:q_linear_deltap}
\end{equation}
where $H$ is a normalized lower-order mean of the fluctuation magnitudes:

\begin{equation}
H(\Delta p)
= \frac{\left(\overline{(\sqrt{|\Delta p'|})^{-1}}\right)^{-1}}{\sqrt{|\overline{\Delta p}|}}
\label{eq:RH_p}
\end{equation}
If the flow fluctuations are instead symmetric, expanding in $q$ gives the ventilation form (\appref{app::FluctDominated}):

\begin{equation}
H(q)
= \frac{\overline{|q'|}}{|q_{\text{MP}}|}
\label{eq:RH_q}
\end{equation}
In both cases, $H$ is a lower-order mean than the root-mean-square that defines $I$: the harmonic mean of $\sqrt{|\Delta p'|}$ for the pressure form and the arithmetic mean of $|q'|$ for the ventilation form. Because the orifice relation is nonlinear, $\Delta p'$ and $q'$ cannot both be symmetric, so the two forms apply to different flows. Real openings need not satisfy either premise exactly, since the fluctuations may be asymmetric in both variables or the flow may not respond instantaneously.

Because power means increase with order, $H$ is bounded above by $H_{\mathrm{U}}$ for both forms (\appref{app::FluctDominated}):

\begin{equation}
H(q) 
\le H_{\mathrm{U}}(q) = \frac{\sqrt{\overline{q'^2}}}{|q_{\text{MP}}|} = I(q)%
\label{eq:RH_bound_q}
\end{equation}

\begin{equation}
H(\Delta p)
\leq H_{\mathrm{U}}(\Delta p) =  \sqrt{\frac{\sqrt{\overline{(\Delta p')^2}}}{|\overline{\Delta p}|}} = \sqrt{2I(\Delta p)}%
\label{eq:RH_bound_p}
\end{equation}

\subsubsection{Transition Regime}
\label{sssec::M-F-Transition}
In the transition regime we cannot separate mean and fluctuation effects. This inseparability is because the flow direction depends on both the mean and fluctuating pressures (i.e., $\operatorname{sgn}(\overline{\Delta p} + \Delta p') \neq \operatorname{sgn}(\overline{\Delta p}) \neq \operatorname{sgn}(\Delta p'))$. 

To represent this regime, we write an expression for $H$ that gives a smooth transition between the mean- and fluctuation-dominated regimes. Specifically, we solve for scaled versions of the $H_\mathrm{U}$ upper bounds established in~\eqref{eq:RH_bound_q} and~\eqref{eq:RH_bound_p} that, when substituted for $H$, give single tangential intersections between the mean- and fluctuation-dominated regimes. Because $\overline{q}$ should vary smoothly with $I$, we join the two branches tangentially. Tangency gives a smooth transition without a fitted parameter, though the true transition depends on the full fluctuation distribution. These models for tangent $H$, denoted as $H_{\mathrm{T}}$, are:
\begin{equation}
H_{\mathrm{T}}(q) = I(q)
\label{eq:HT_q}
\end{equation}
and
\begin{equation}
H_{\mathrm{T}}(\Delta p) = \sqrt{\frac{3\sqrt{3}}{8}I(\Delta p)}
\label{eq:HT_p}
\end{equation}
Using $H_{\mathrm{T}}$ gives a model for $\overline{q}$ which accounts for both the fluctuation-dominated and transition regimes:
\begin{equation}
\overline{q} \approx \frac{q_{\text{MP}}}{2H_{\mathrm{T}}}
\label{eq:transition}
\end{equation}

\subsubsection{Non-dimensional Combined Piecewise Model}
\label{sssec::M-F-Piecewise}

The final combined model is written in non-dimensional form using non-dimensional ventilation rates and pressure differences $q_n=q/(U_{10}A)$ and $\Delta C_p=\Delta p/(\frac{1}{2}\rho U_{10}^2)$, where $A$ is the opening area and $U_{10}$ is the prescribed wind velocity at 10 m height. The fluctuation corrections can be expressed equivalently in terms of these nondimensional quantities, replacing $\sigma_q$ and $\sigma_{\Delta p}$ with $\sigma_{q_n}$ and $\sigma_{\Delta C_p}$, respectively. We refer to these nondimensional fluctuation intensities collectively as $\sigma_n$.

In this non-dimensional form, we can combine~\eqref{eq:transition} with the mean-dominated expression~\eqref{eq:qAFN_qstd} to obtain a smooth piecewise function dependent only on $I$:

\begin{equation}
\overline{q}_n \approx q_{\mathrm{PW},n}
= q_{\mathrm{MP},n} \cdot F_{\mathrm{fc}}(I)
\label{eq:piecewise_model}
\end{equation}

\begin{equation}
F_{\mathrm{fc}}(I) =
\begin{cases}
\dfrac{1}{2H_{\mathrm{T}}} & I \geq I^*, \\[6pt]
\sqrt{1 - I^2} & I < I^*.
\end{cases}
\label{eq:fc_factor}
\end{equation}

where $q_{\text{PW},n}$ is the non-dimensionalized ventilation rate from the piecewise model and $F_{fc}$ is the fluctuation correction factor. $I^*$ is the transition point:
$I^*(\Delta p) = 1/\sqrt{3}$ and $I^*(q) = 1/\sqrt{2}$. $F_{fc}$ can span from near 0 (a large correction) to 1 (no correction). In the absence of fluctuations, $F_{fc}(0) = 1$, and this model reduces to the traditional mean-pressure formulation.

\subsection{Verification of Fluctuation Corrections}
\label{ssec::M-MonteCarlo}

Here we aim to verify that the proposed fluctuation corrections address the square-root nonlinearity in the orifice equation. We perform this verification by generating synthetic fluctuation data and comparing the fluctuation-corrected predictions of the mean ventilation rate with corresponding sample-based estimates.

\subsubsection{Synthetic data generation}
\label{sssec::M-MC-Setup}

We generated two synthetic ventilation datasets by sampling either
non-dimensional ventilation rates or pressure coefficient differences from normal distributions.
Each dataset contains random series that span 
fluctuation-dominated to mean-dominated flow by varying the mean value $\mu$ while holding the standard deviation $\sigma$ fixed. We used 40 logarithmically spaced mean values across the prescribed range.

For the ventilation-based correction, we sample non-dimensional ventilation rates from normal distributions with means $\mu=\overline{q_n}$ ranging from 0.01 to 3 and a fixed standard deviation $\sigma_{q_n}=1$. For each mean value, we draw $10^8$ samples and calculate their arithmetic mean as the reference ventilation rate. We also convert each sample to a corresponding pressure difference using the inverse orifice equation. Applying the orifice equation to the mean of these pressure differences gives the mean-pressure prediction ($q_{\mathrm{MP},n}$). Finally, we evaluate $I(q_n)$ (Eq.~\ref{eq:FQ_qstd}), $H(q_n)$ (Eq.~\ref{eq:RH_q}), $H_U(q_n)$ (Eq.~\ref{eq:RH_bound_q}), and $H_T(q_n)$ (Eq.~\ref{eq:HT_q}) to calculate the different fluctuation-corrected predictions.

For the pressure-based correction, we sample nondimensional pressure coefficient differences from normal distributions with means $\mu=\overline{\Delta C_p}$ ranging from 0.0536 to 24.11 and a fixed standard deviation $\sigma_{\Delta C_p} = 2(\sigma_{q_n}/C_d)^2 = 5.357$, with $C_d = 0.611$. These values give approximately the same range of $q_{\mathrm{MP},n}$ and the same fluctuation scale as the ventilation-based experiment. For each mean value, we draw $10^8$ pressure difference samples and convert each to a corresponding ventilation rate using the orifice equation. The arithmetic mean of these ventilation rates provides the reference ventilation rate. We obtain the mean-pressure prediction $q_{\mathrm{MP},n}$ by applying the orifice equation to the mean of the sampled pressure differences. Finally, we evaluate $I(\Delta C_p)$ (Eq.~\ref{eq:FQ_pstd}), $H(\Delta C_p)$ (Eq.~\ref{eq:RH_p}), $H_U(\Delta C_p)$ (Eq.~\ref{eq:RH_bound_p}), and $H_T(\Delta C_p)$ (Eq.~\ref{eq:HT_p}) to calculate the different fluctuation-corrected predictions.

\subsubsection{Verification Results}
\label{sssec::M-MC-Results}
\begin{figure}[htb]
    \centering
    \includegraphics[width=0.99\textwidth]{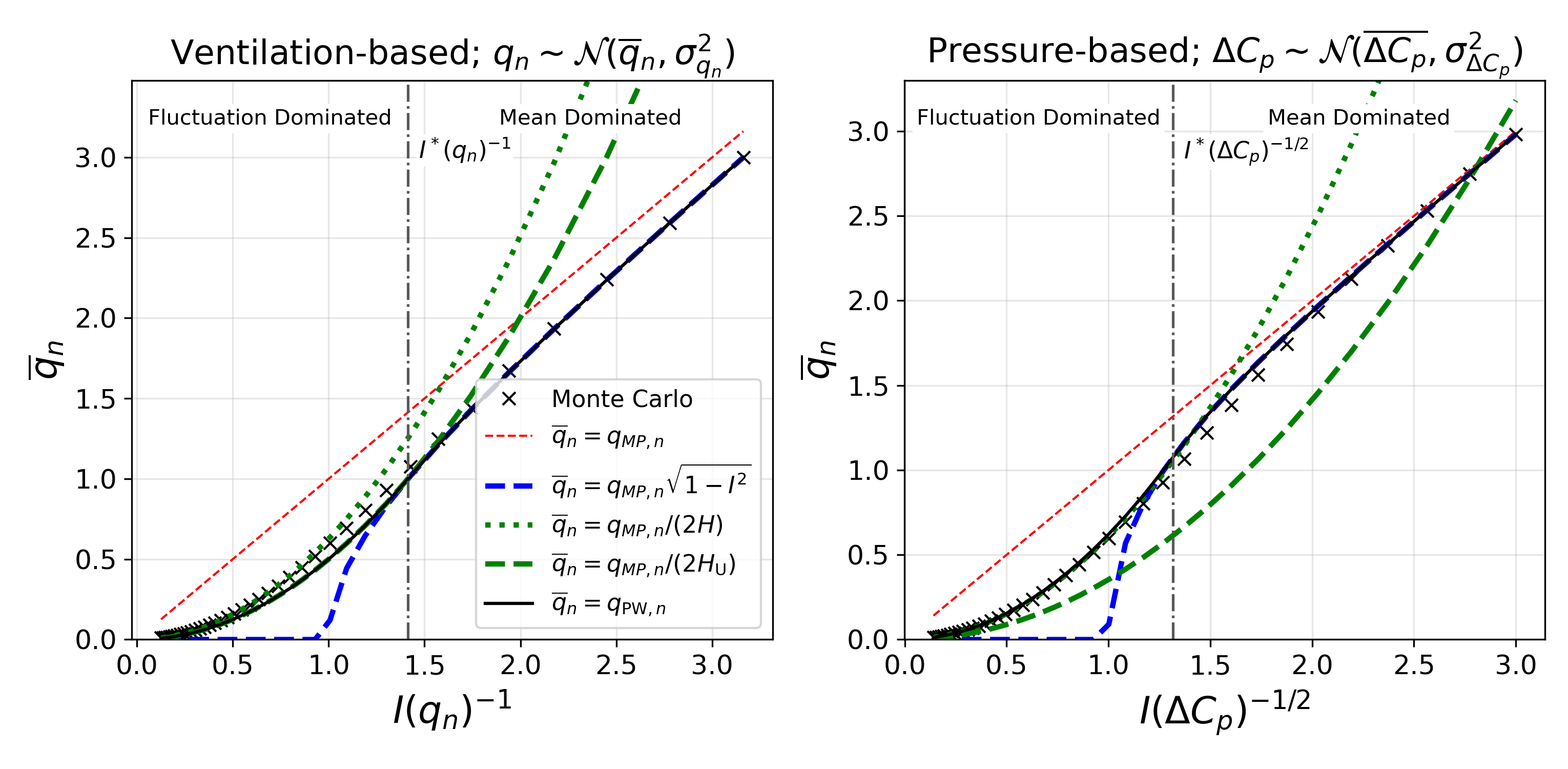}
    \caption{Monte Carlo analysis comparing the mean-pressure ventilation predictions against the actual quasi-steady ventilation rates. Lines show the mean-dominated, fluctuation-dominated, and piecewise models. The left column shows results from the ventilation formulation, while the right column shows results from the pressure formulation. } 
    \label{fig:monteCarlo}
\end{figure}

Figure~\ref{fig:monteCarlo} plots the verification results for the fluctuation corrections based on the ventilation rate (left) and the pressure (right), with the mean ventilation rate $\overline{q}_n$ on the y-axis, and $I(q)^{-1}$ or $I(\Delta p)^{-1/2}$ on the x-axis. This representation produces a 1:1 line for the mean-pressure value $q_{\mathrm{MP},n}$. In addition, the plot shows the sample mean, and the mean-dominated, fluctuation-dominated, and piecewise models proposed in Section~\ref{ssec::M-Corrections}. %

In the mean-dominated regime ($I^{-1} > (I^*)^{-1}$), both formulations agree closely with the true sample mean. In the fluctuation-dominated regime ($I^{-1} < (I^*)^{-1}$), models using $H$ match the true sample mean for both formulations, while the bounds $H_{\mathrm{U}}$ underpredict it. For Gaussian fluctuations, $H_{\mathrm{U}}$ exceeds the exact $H$ by a constant factor. The piecewise model ($q_\text{PW}$), which uses $H_{\mathrm{T}}$ to bridge the two regimes, closely follows the true sample mean across all $I^{-1}$. \appref{app::GaussianH} analytically quantifies how closely these models match the exact Gaussian fluctuation results in the fluctuation-dominated regime. %

\subsection{Fluctuation-Corrected Airflow Network Model Implementation}
\label{ssec::M-CorrectedAFN}

In this section we discuss modeling choices and practical considerations for integrating fluctuation corrections into airflow networks. 

\subsubsection{Specifying Fluctuation Intensities}
\label{sssec::M-F-P-VvsP}

\begin{table}[h]
\centering
\begin{tabular}{lcccc}
\hline
& \multicolumn{2}{c}{Windows}
& \multicolumn{2}{c}{Skylights} \\
 Bulk $\sigma_n$ & in (+) & out (-) & in (+) & out (-) \\
\hline
$\sigma_{q_n}$
& 0.12 & 0.13 & 0.13 & 0.12 \\
$\sigma_{\Delta C_p}$
& 0.12 & 0.13 & 0.13 & 0.13 \\
\hline
\end{tabular}
\caption{Bulk $\sigma$ values measured from the LESs.}
\label{table:sigmaParams}
\end{table}

The fluctuation corrections require specifying the fluctuation intensities with either $\sigma_{q_n}$ or $\sigma_{\Delta C_p}$. These $\sigma$ can be specified separately for each opening or represented as bulk values for groups of openings with similar characteristics. To evaluate these options, we consider both opening-specific and bulk intensities for the ventilation-based and pressure-based corrections, using values extracted from the available LES datasets. The values for $\sigma_{q_n}$ were determined from the coupled indoor-outdoor LESs, while the values for $\sigma_{\Delta C_p}$ were obtained from the exterior-only LESs. For the bulk representation, we average $\sigma$ across all simulated cases and wind directions, separately for four categories: windows that act as an inflow, windows that act as an outflow, skylights that act as an inflow, and skylights that act as an outflow. Table \ref{table:sigmaParams} shows these bulk $\sigma_n$ values. %

\subsubsection{In-network or post-network implementation}
\label{sssec::M-F-P-AFN}

The fluctuation corrections can be implemented in an airflow network using two approaches. The first, which we term in-network, is an intrusive approach that incorporates the fluctuation corrections directly into the airflow network equations. The network solves for the internal pressures such that the fluctuation-corrected ventilation rates satisfy continuity at each interior node. This is the more physically consistent approach, as the corrected ventilation rates are used directly in the mass balance. However, implementing it in an existing airflow network solver requires modifying the equations used to determine the internal pressures.

The second, which we term post-network, is a non-intrusive approach that applies the fluctuation corrections after solving the standard airflow network. The network is first solved using the mean-pressure formulation with equal $C_d$ at each opening. This solve gives the internal pressures and corresponding pressure differences across each opening. These pressure differences are then used to calculate ventilation rates using the fluctuation-corrected models. This approach is easier to implement in existing airflow network models, but the corrected ventilation rates do not necessarily satisfy continuity, since the internal pressures are not updated to account for the corrections.

\subsection{Summary of Models Evaluated and $C_d$ Calibration}
\label{ssec::M-Evaluated}

\subsubsection{Summary of Models Evaluated}
The implementation choices described in Section~\ref{ssec::M-CorrectedAFN} lead to many possible airflow network setups. We evaluated a wide range of possible combinations, but we restrict the results discussed in this paper to the model choices listed in Table~\ref{table:fitParams}. We selected these formulations because they demonstrate the effects of using geometry-specific pressure coefficients and implementing fluctuation corrections separately. For the fluctuation corrections, we consider implementations that span the different levels of complexity using both $\sigma_q$ or $\sigma_{\Delta p}$. Specifically, we will present corrections based on opening-specific $\sigma$ values implemented in-network (IN), corrections based on bulk $\sigma$ values implemented in-network (IN),  and corrections based on bulk $\sigma$ values implemented post-network (PN).

\subsubsection{Calibration of $C_d$}
\label{sssec::M-Calibration}

Airflow network models require specifying a value for the discharge coefficient $C_d$. For square, sharp-edged openings, the still-air value is $C_d = 0.611$, but it has been shown that $C_d$ values under real flow conditions can be different~\citep{karava_wind_2004, shirzadi_development_2018, hwang_largeeddy_2022}. To assess the performance of adding fluctuation-corrections to the standard model form without relying on an idealized $C_d$ value, we use the ventilation rates observed in the coupled exterior-interior LESs to determine the values of $C_d$ that minimize the root mean square error (RMSE) between the fluctuation-corrected airflow network predictions and the LES observations. Consistent with the bulk fluctuation intensities, separate coefficients are estimated for four opening categories: inflow windows, outflow windows, inflow skylights, and outflow skylights. We do not modify $C_d$ for the interior door in the dual-room configuration (with a three times larger area than the windows), maintaining $C_d = 0.611$ and no fluctuation correction.

Table \ref{table:fitParams} reports the $C_d$ values used in the different airflow network models. 
For the ASHRAE mean-pressure formulation, the calibration process attempts to compensate for the significant prediction error introduced by the ASHRAE pressure coefficients and produces non-physical $C_d$ values. Hence, we evaluated the ASHRAE Mean-Pressure Baseline model using a standard $C_d$ of 0.611. To assess the improvements that can be obtained by specifying more accurate pressure coefficients without compounding effects of changing $C_d$, we also present the values for the LES Mean-Pressure Baseline using a $C_d$ value of 0.611. We note that for this model, the fitting produced $C_d$ values between 0.58 and 0.60. This similarity to the potential-flow-derived baseline of $0.611$ is coincidental and dependent on fluctuation errors.

For the fluctuation-corrected models, the values shown for the post-network implementation were estimated by nonlinear least squares (SciPy \texttt{curve\_fit}{}~\cite{virtanen_scipy_2020}). For $\sigma_{\Delta C_p}$ the prediction is linear in $C_d$, so this reduces to least squares through the origin. For the in-network implementation, calibration was performed by repeatedly solving the airflow network as the coefficients are varied. We use the gradient-free Nelder--Mead optimization algorithm implemented in SciPy \citep{virtanen_scipy_2020, gao_implementing_2012}, initialized with the post-network bulk $C_d$ estimates. %
The resulting $C_d$ values reported in Table \ref{table:fitParams} are generally similar for the different fluctuation-corrected models. %

\begin{table}[h]
\centering
\scriptsize
\setlength{\tabcolsep}{4pt}
\begin{tabular}{llcccc}
\hline
& & \multicolumn{2}{c}{\textbf{Windows}}
    & \multicolumn{2}{c}{\textbf{Skylights}} \\
\textbf{Pressure source} & \textbf{Fluctuation correction}
    & in (+) & out (-) & in (+) & out (-) \\
\hline
ASHRAE & Mean-Pressure Baseline
    & 0.611& 0.611& -- & -- \\
\hline
LES & Mean-Pressure Baseline
    & 0.611& 0.611& 0.611& 0.611\\
\hline
LES & $\sigma_{q_n}$ IN Opening-Specific
    & 0.67 & 0.71 & 0.67 & 0.73 \\
LES & $\sigma_{q_n}$ IN Bulk
    & 0.66 & 0.72 & 0.66 & 0.72 \\
LES & $\sigma_{q_n}$ PN Bulk
    & 0.69 & 0.69 & 0.70 & 0.69 \\
\hline
LES & $\sigma_{\Delta C_p}$ IN Opening-Specific
    & 0.65 & 0.72 & 0.65 & 0.76 \\
LES & $\sigma_{\Delta C_p}$ IN Bulk
    & 0.65 & 0.69 & 0.64 & 0.73 \\
LES & $\sigma_{\Delta C_p}$ PN Bulk
    & 0.67 & 0.67 & 0.68 & 0.71 \\
\hline
\end{tabular}
\caption{Eight reported airflow networks and their $C_d$ values.}
\label{table:fitParams}
\end{table}

\section{Analysis of Airflow Network Model Improvements}
\label{sec::Results}

This section evaluates the improvements in natural ventilation predictions obtained by using geometry-specific surface pressures and fluctuation corrections in airflow network models. Model predictions are compared against ventilation rates obtained from the coupled interior-exterior LESs. We first assess the overall model performance using RMSE and bias, then examine the contributions of geometry-specific surface pressures and fluctuation corrections separately.

\subsection{Overview of Model Performance}

\begin{figure}[htb]
    \centering
    \includegraphics[width=0.9\textwidth]{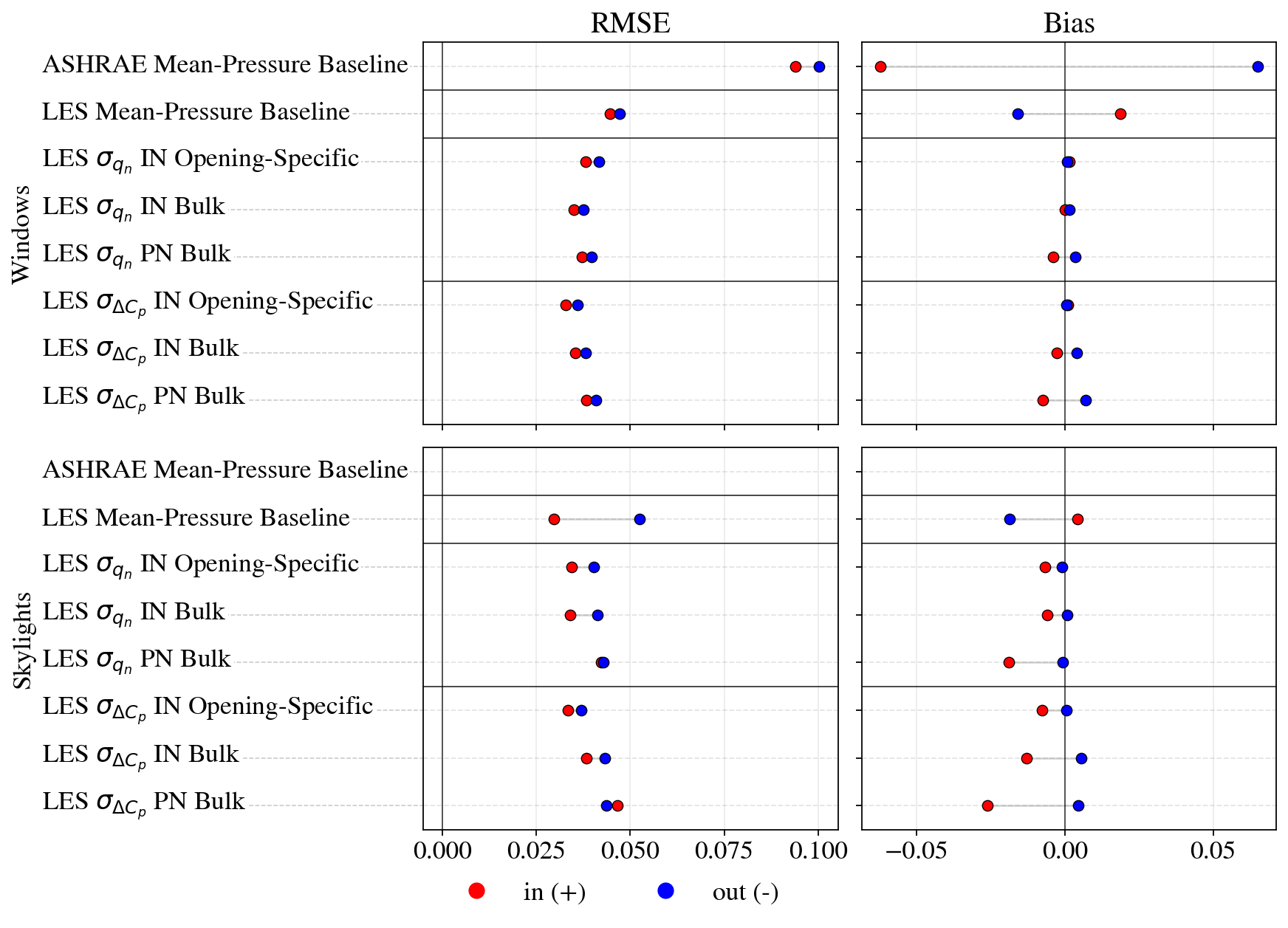}
    \caption{RMSE and bias for flow through openings across airflow network models. Top row shows windows while bottom shows skylights. Horizontal lines match Table~\ref{table:fitParams} by dividing the results by pressure source and fluctuation correction type.} 
    \label{fig:Error_Windows}
\end{figure}

\begin{figure}[htb]
    \centering
    \includegraphics[width=0.9\textwidth]{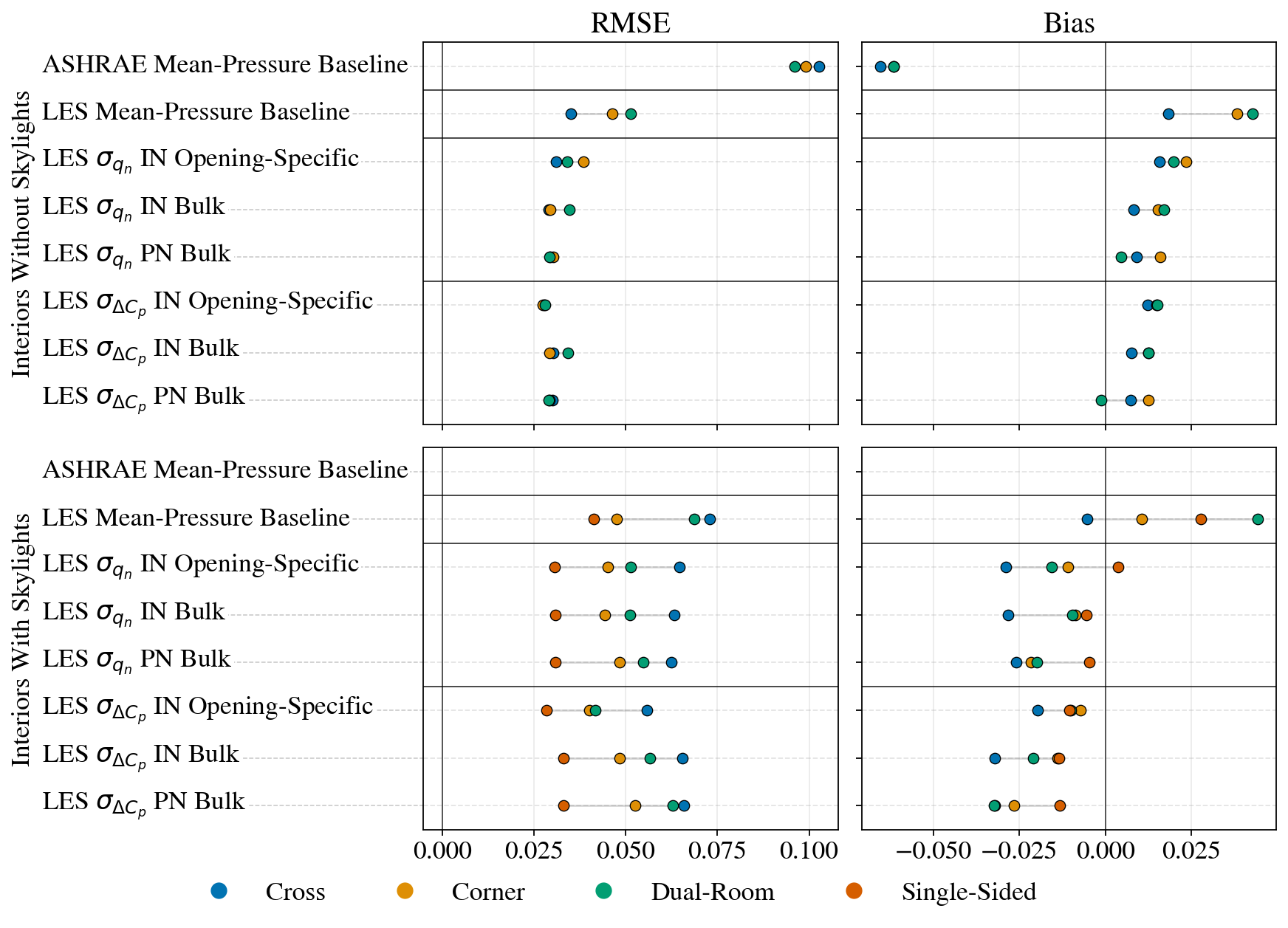}
    \caption{RMSE and bias for flow through interior sections across airflow network models. Top row shows rooms without skylights while bottom row shows rooms with skylights. } 
    \label{fig:Error_Rooms}
\end{figure}

 Figure~\ref{fig:Error_Windows} presents the RMSE and bias for the predicted ventilation rates through individual windows and skylights, while Figure~\ref{fig:Error_Rooms} shows these error statistics for the total ventilation rates across the four interior configurations: cross, corner, dual-room, and single-sided. The latter comparison allows us to assess model performance across different ventilation configurations and evaluate the accuracy of the predicted total ventilation rates. Together, these figures indicate that the use of geometry-specific pressures provides the largest improvement in predictive performance. The fluctuation corrections further refine the model, generally leading to additional reductions in the RMSE and bias. The following sections examine the improvements associated with geometry-specific surface pressures and fluctuation corrections in more detail. %
 
\subsection{Improvement from Geometry-Specific Surface Pressures}
\label{ssec::R-GeoPressures}

\begin{figure}[htb]
    \centering
    \includegraphics[width=0.9\textwidth]{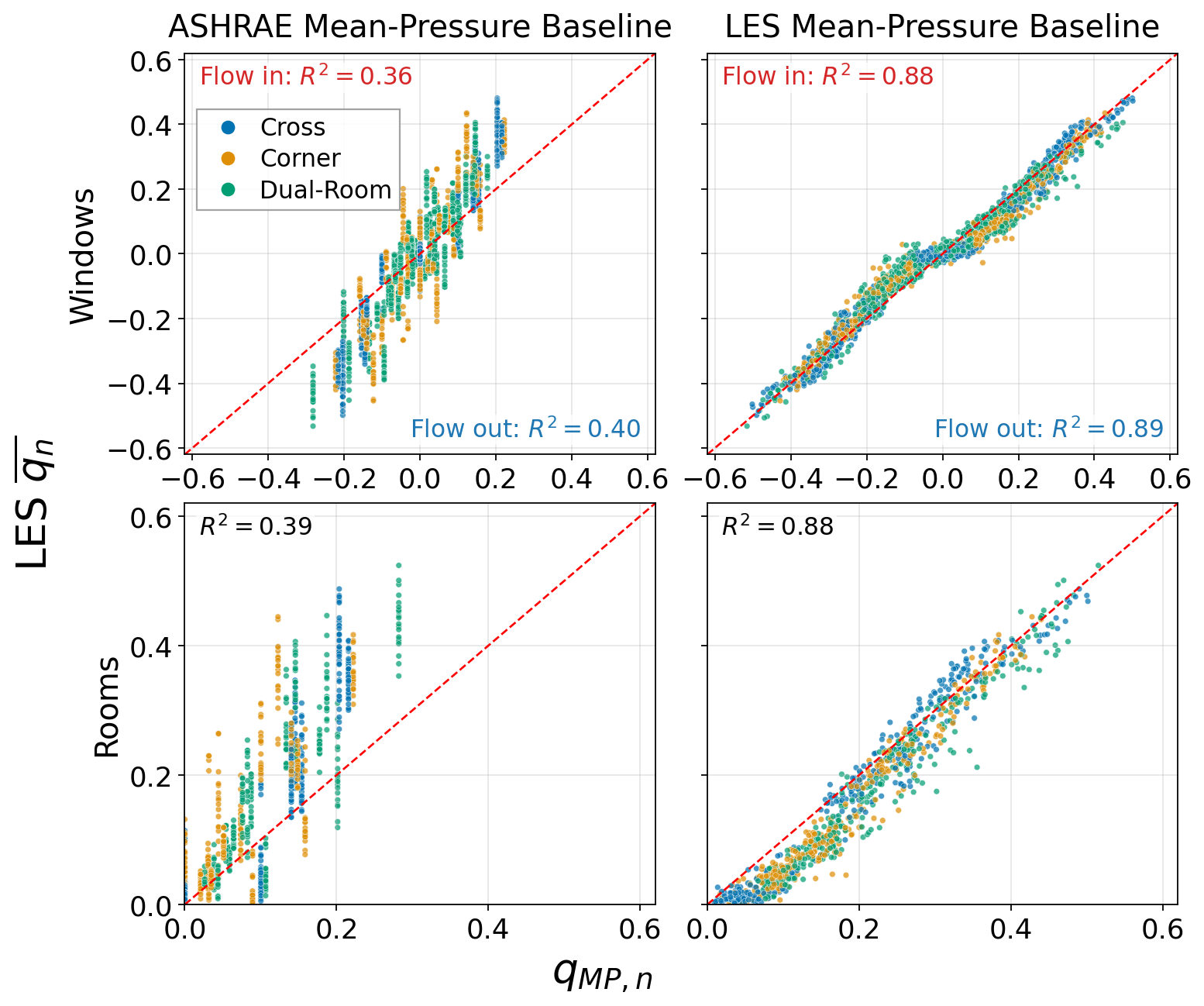}
    \caption{Baseline mean-pressure airflow network predictions versus ventilation rates from coupled interior-exterior LESs. Airflow network predictions in the left column use ASHRAE pressures and in the right column use geometry-specific pressures. The red dashed line denotes perfect model fit. All results are from interiors without skylights.}
    \label{fig:ASHRAE}
\end{figure}

For windows in houses without skylights, which the ASHRAE baseline covers, replacing ASHRAE surface pressures with geometry-specific LES pressures reduces the RMSE of predicted ventilation rates by 56--58\% and the bias by approximately 55\%. Across all openings and interiors in Figures~\ref{fig:Error_Windows} and \ref{fig:Error_Rooms}, the RMSE reduction is about 50\% and the bias reduction ranges from about 30\% (dual-room interiors) to 75\% (outflow windows). To examine how these improvements vary across individual cases, Figure~\ref{fig:ASHRAE} shows scatter plots comparing the ventilation rates predicted by the two mean-pressure airflow network models against those obtained from the coupled interior-exterior LESs.

The scatter plots confirm that geometry-specific pressures substantially improve agreement with the LES results, increasing the $R^2$ from 0.36 to 0.88 for inflow windows, from 0.40 to 0.89 for outflow windows, and from 0.39 to 0.88 for interiors. The ASHRAE-based model predicts identical ventilation rates for several cases with different LES ventilation rates. This discrepancy occurs because the standardized ASHRAE pressure coefficients do not capture differences in the local pressure distributions caused by the surrounding building geometry, resulting in identical pressure inputs for cases that experience different flow conditions. The geometry-specific pressures capture these differences, leading to a much better representation of the simulated ventilation rates.

Despite these significant improvements, the scatter plots still show systematic nonlinear deviations from the 1:1 line. These remaining errors motivate the fluctuation corrections examined next.

\subsection{Improvement from Fluctuation-Corrections}
\label{ssec::R-FluctuationCorrections}

We first quantify how much fluctuation corrections further reduce model error (Section~\ref{sssec::R-F-Effect}). We then discuss the effects of implementation simplifications (Section~\ref{sssec::R-F-Validating}).

\subsubsection{Overall Effect}
\label{sssec::R-F-Effect}

Building on the improvements obtained with geometry-specific surface pressures, the fluctuation corrections generally further reduce RMSE and bias, although their effectiveness varies across correction methods and opening types. As shown in Figure~\ref{fig:Error_Windows}, the corrections improve the RMSE for inflow and outflow through windows by approximately 10--25\%. Similar improvements are observed for outflow through skylights, which is the more common flow direction due to buoyancy, while predictions for skylight inflows generally become less accurate. At the interior level (Figure~\ref{fig:Error_Rooms}), the corrections also typically reduce the RMSE by approximately 10--45\%.

\begin{figure}[htb]
    \centering
    \includegraphics[width=0.9\textwidth]{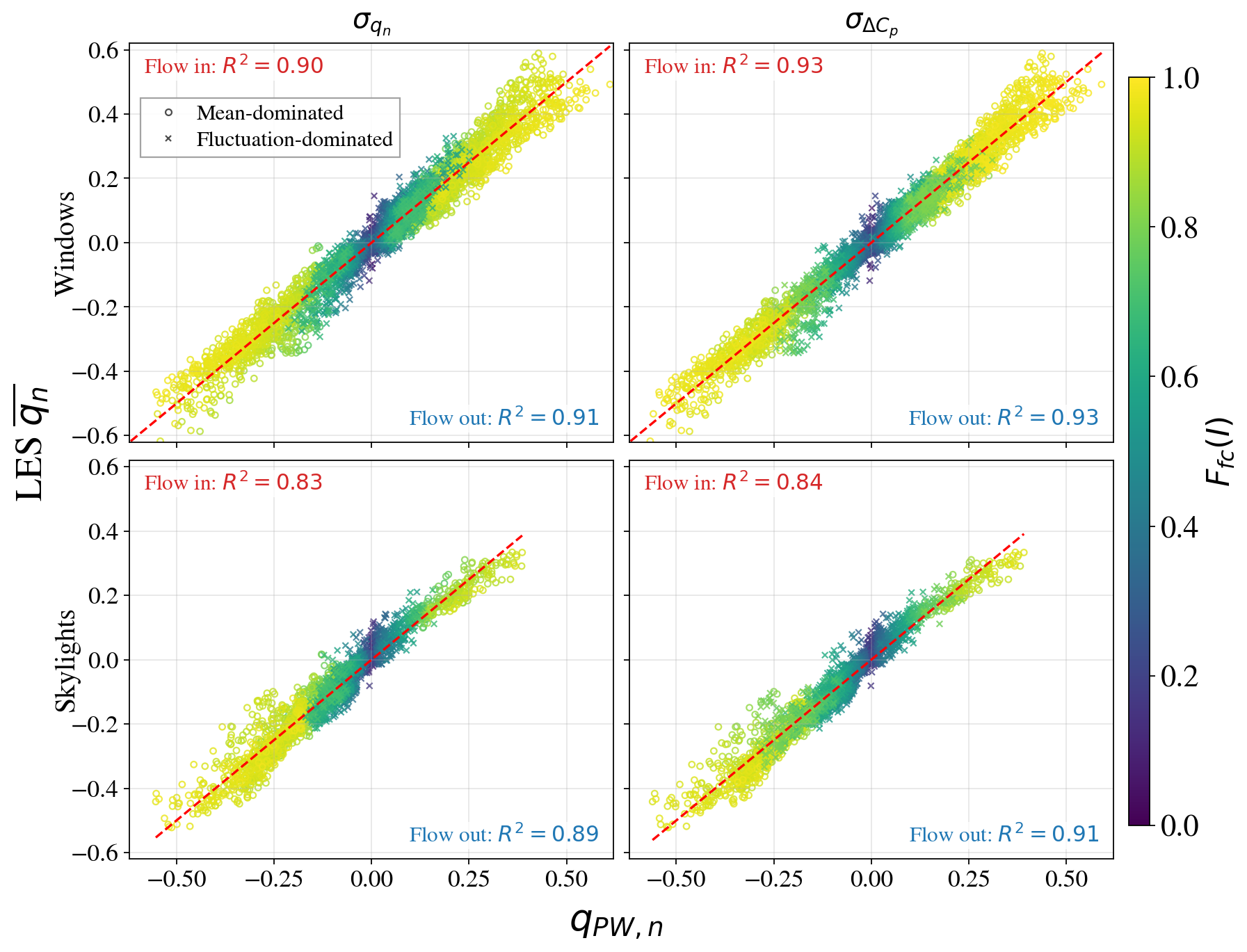}
    \caption{Opening-specific fluctuation-corrected ventilation rates versus measured ventilation rates from coupled interior-exterior LESs. Colors show the magnitude of the fluctuation correction. The dashed red line denotes perfect model fit.}
    \label{fig:I_pointwise}
\end{figure}

Figure \ref{fig:I_pointwise} shows the modeled ventilation rates after applying in-network opening-specific fluctuation corrections. The fluctuation corrections appear to eliminate most of the nonlinearity between the measured and mean-pressure ventilation rates shown in Figure~\ref{fig:ASHRAE}. The points are colored by the magnitude of the fluctuation-correction factor $F_{fc}$ \eqref{eq:fc_factor}, showing that $F_{fc}$, despite being determined using the opening-specific $\sigma_{n}$, mostly scales with $q_{PW,n}$. This scaling provides a first indication that a fluctuation-correction determined using a bulk value of $\sigma_n$ could also provide accurate results.

\subsubsection{Effect of Implementation Simplifications}
\label{sssec::R-F-Validating}

\paragraph{Bulk vs Opening-Specific Fluctuation Corrections}
\begin{figure}[htb]
    \centering
    \includegraphics[width=0.9\textwidth]{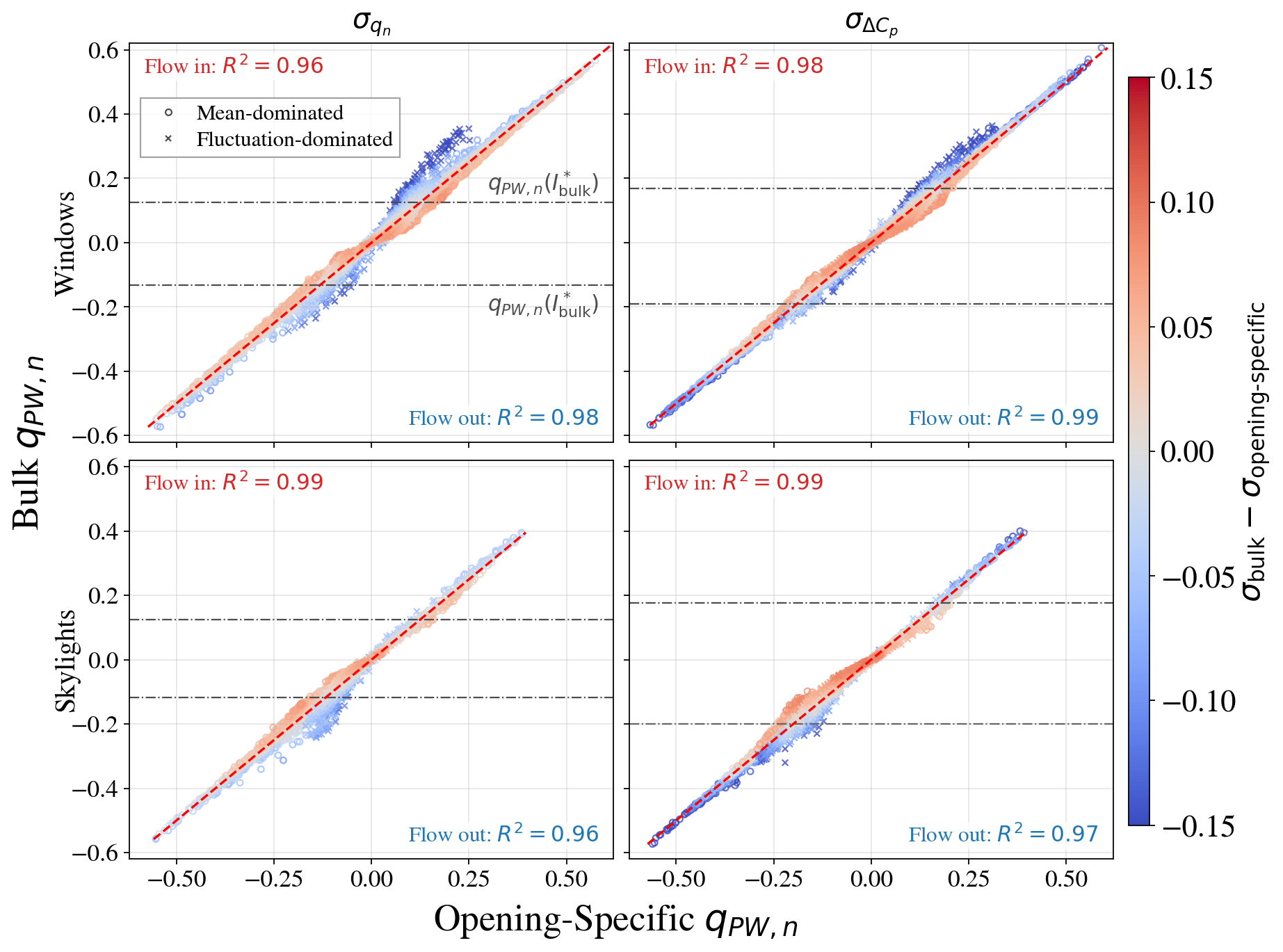}
    \caption{Comparing fluctuation-corrected models using the opening-specific and bulk $\sigma$. Colors show the difference between the bulk and opening specific $\sigma_{n}$ values. The horizontal lines denote the transitions between the mean- and fluctuation-dominated regimes for the bulk model, while the point styles show the mean- and fluctuation-dominated regimes for the opening-specific model. Both models use the in-network implementation. The dashed red line denotes perfect model fit.}
    \label{fig:BvOS}
\end{figure}

Figure~\ref{fig:BvOS} directly compares the predictions obtained using the bulk vs opening-specific fluctuation corrections. Generally, the predicted ventilation rates are very close, with $R^2$ ranging from 0.96 to 0.99. If the bulk $\sigma$ is smaller than the opening specific $\sigma$, it will always over predict the ventilation rate, and vice versa. The largest discrepancies occur when the bulk $\sigma_n$ is about 0.1--0.3 smaller than the opening-specific $\sigma_n$. In these cases, the opening-specific model is often in the fluctuation-dominated regime, while the bulk-model is in the mean-dominated regime. However, these larger discrepancies do not occur frequently, and the aggregate difference in model error between the opening-specific and bulk models, reported in Figures~\ref{fig:Error_Windows} and \ref{fig:Error_Rooms}, is small.  %

\paragraph{Post-Network vs In-Network Implementation}

\begin{figure}[htb]
    \centering
    \includegraphics[width=0.9\textwidth]{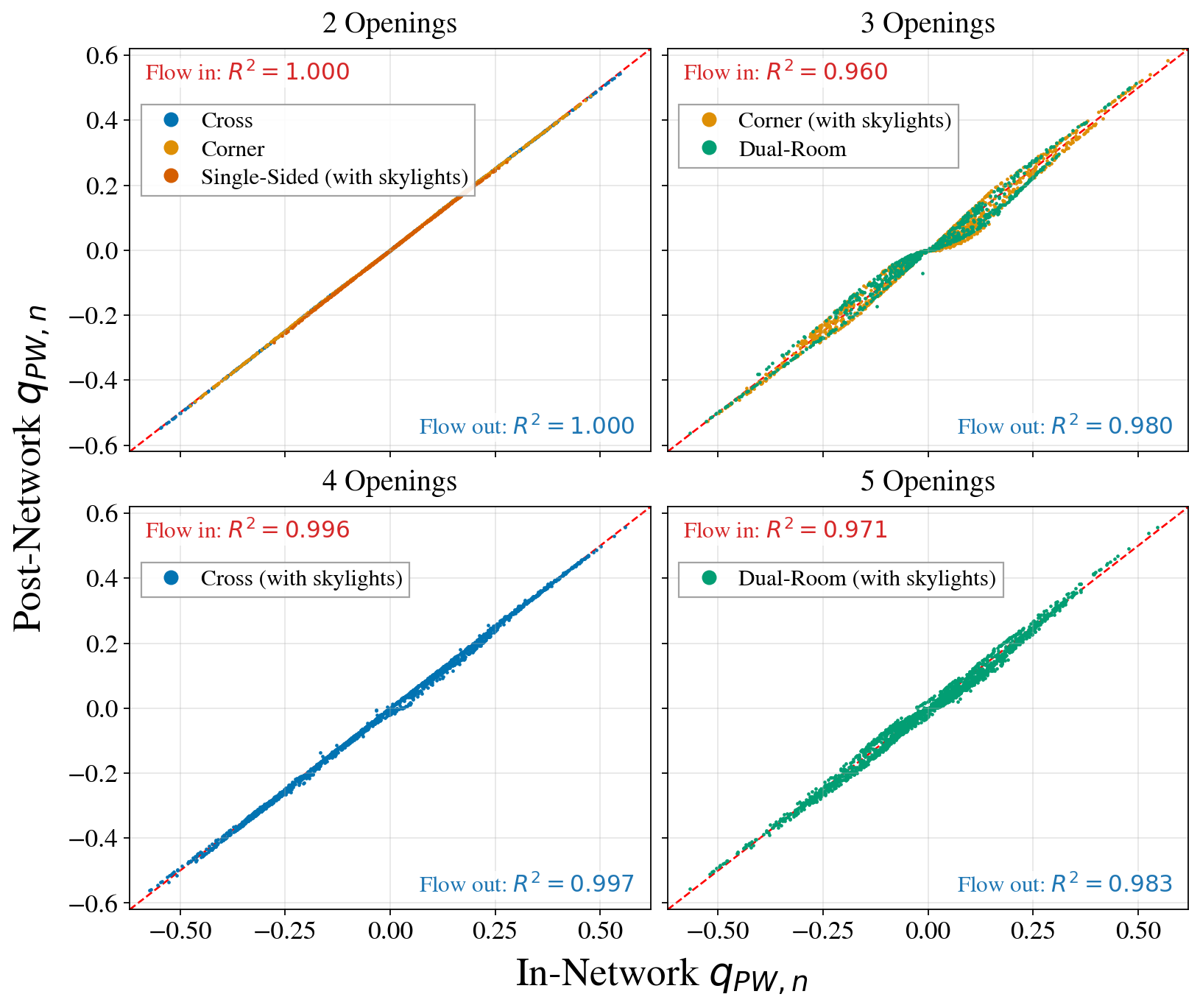}
    \caption{In-Network vs Post-Network fluctuation corrections using the bulk $\sigma_{\Delta C_p}$.}
    \label{fig:InvPV}
\end{figure}

Figure~\ref{fig:InvPV} compares the in-network and post-network model performance using bulk $\sigma_{\Delta C_p}$. We present the results separately for rooms with different numbers of openings, since we expect the main error arises from not exactly satisfying mass balance in the post-network implementation. Despite this drawback, the model agreement is generally very good, with the lowest $R^2$ value equal to 0.96. The cases with two openings have near perfect agreement, because the implementations would be identical for identical $C_d$ and $\sigma$ values, and the actual values used (Tables~\ref{table:sigmaParams} and \ref{table:fitParams}) are generally very close. On the other hand, the three-opening case shows the most sensitivity to the implementation method. With three openings, the ventilation flow rate through two openings must sum to the third, creating asymmetric ventilation rates between the inflows and outflows. Since the fluctuation correction factor $F_{fc}$ is strongly tied to the modeled ventilation rate, the inflow-outflow asymmetry leads to asymmetric fluctuation corrections. The in-network method considers this asymmetry when solving for the interior pressure $\Delta p_0$, while the post-network method does not. The four- and five-opening cases tell a similar story, with the more symmetric four opening case showing better agreement than the asymmetric five opening case. Compared to in-network, the post-network implementation performs slightly worse for openings (Figure~\ref{fig:Error_Windows}) but similar or slightly better for interiors (Figure~\ref{fig:Error_Rooms}).

\paragraph{Bulk Post-Network Model Performance}

\begin{figure}[htbp]
    \centering
    \includegraphics[width=0.99\textwidth]{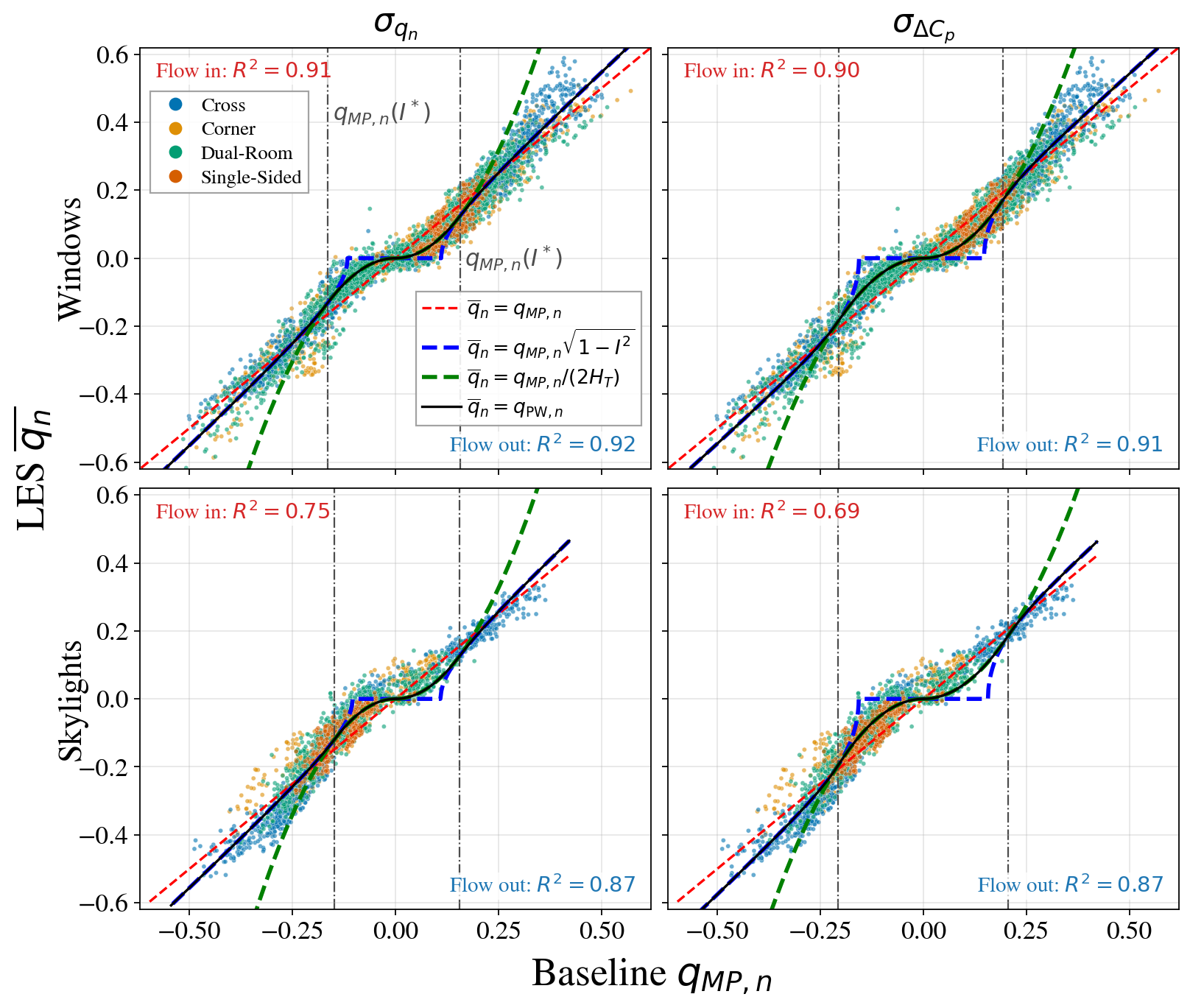}
    \caption{Bulk post-network models overlaid on ventilation results. The x-axis shows baseline mean-pressure ventilation rates using geometry-specific pressures while the y-axis shows ventilation rates from coupled interior-exterior LESs. Lines show the mean-pressure prediction, mean-dominated regime, fluctuation-dominated regime, and piecewise model.}
    \label{fig:I_bulk}
\end{figure}

To conclude our analysis of the fluctuation-correction implementations, Figure~\ref{fig:I_bulk} quantifies the performance of the simplest implementation, i.e., the bulk post-network models. The scatterplots compare the mean-pressure ventilation rate against the measured ventilation rates. Over this data, the solid black curves plot the fluctuation-corrected models~\eqref{eq:piecewise_model}, demonstrating that the models follow the nonlinearity in the data with a trend that is similar to the Monte Carlo results in Figure~\ref{fig:monteCarlo}.

\section{Discussion}
\label{sec::Discussion}

The results demonstrate that geometry-specific surface pressures and fluctuation corrections can substantially improve airflow network predictions of natural ventilation. In this section, we discuss the implications for selecting practical fluctuation correction formulations (Section~\ref{ssec::D-Preferred}) and specifying discharge coefficients (Section~\ref{ssec::D-Coefficients}).

\subsection{Recommended Formulations}
\label{ssec::D-Preferred}

We recommend pressure-based fluctuation corrections. $\sigma_{\Delta C_p}$ can be obtained from simulations or measurements of the exterior flow field alone, while $\sigma_{q_n}$ requires ventilation rate information from coupled interior-exterior flow. Models using $\sigma_{\Delta C_p}$ and $\sigma_{q_n}$ generally perform similarly.

Using opening-specific $\sigma_{\Delta C_p}$ offers marginally better performance than using a bulk value. If opening-specific values are available, we recommend using them. However, the relatively small difference in performance suggests that bulk values provide a practical alternative when detailed fluctuation information is unavailable.  Developing estimates of bulk values as a function of overall canopy geometry and wind conditions would be a substantially less demanding modeling task than estimating opening-specific values.

Similarly, the in-network pressure models have the benefit of preserving mass conservation and generally perform slightly better than the post-network models at predicting ventilation through openings (but slightly worse for interiors). The in-network implementation requires modifying the network equations, but its computational complexity is similar to the standard approach, which typically already requires an iterative solution for $\Delta p_0$. We therefore recommend implementing the in-network formulation when feasible, particularly for configurations where the fluctuation intensities are significantly different at the inflow and outflow. In these cases post-network corrections can introduce significant mass imbalances. When modifying the solver is not practical, bulk pressure-based post-network corrections offer a simpler alternative that is likely sufficient for most applications.

\subsection{Selection of Discharge Coefficients}
\label{ssec::D-Coefficients}

\subsubsection{Recommended Values and Remaining Uncertainty}
With geometry-specific surface pressures, the least-squares calibration gives discharge coefficients of 0.58--0.60 for the LES mean-pressure model (Section~\ref{sssec::M-Calibration}) and approximately 0.7 for the fluctuation-corrected models (Table~\ref{table:fitParams}). 
The values for the fluctuation-corrected models are higher than the potential-flow-derived baseline of $0.611$, but agree well with the 0.6--0.8 range reported in past experimental and simulation results for cases without cross-flow~\citep{etheridge_natural_2012, chiu_external_2007, costola_unsteady_2008}. Based on these results, we recommend using $C_d = 0.7$ in combination with the fluctuation-corrected models for building geometries and sharp-edged square openings similar to those investigated here. However, the extent to which this recommendation transfers to other floor plans, ventilation configurations, and urban canopy geometries requires further investigation.

To assess whether uncertainty in the fitted parameters could explain the remaining prediction errors, we also performed a Bayesian calibration of the bulk post-network models, treating $C_d$, the bulk $\sigma_n$, and residual error as uncertain quantities. The resulting posterior distributions of $C_d$ and the fluctuation intensities were relatively narrow, while the estimated residual error standard deviations were comparable to the model RMSE. Within the assumed statistical model, this suggests that uncertainty in the calibrated parameters is not a dominant source of the remaining prediction errors. 

\subsubsection{Inadvertent Compensation for Neglected Fluctuation Effects}

The observation that the least-squares calibration gives discharge coefficients from 0.58--0.60 for the mean-pressure models and 0.64--0.76 for the fluctuation-corrected models indicates that neglecting fluctuations in airflow network models may lead to compensating adjustments in other model parameters. For example, previous studies have reported reductions in $C_d$ associated with cross-flow at openings~\citep{sawachi_wind_2004, chiu_external_2007, costola_unsteady_2008, etheridge_natural_2012}. Because cross-flow can also influence fluctuations, some of these observed changes in effective discharge coefficients may reflect a combination of physical effects rather than changes in the discharge coefficient alone. 

A second example concerns the relationship between ventilation rate and mean pressure difference. The fluctuation-correction formulation reveals different scaling behavior in mean-dominated and fluctuation-dominated regimes. If $\sigma$ is independent of $\Delta p$, the mean-dominated regime gives $q \sim \sqrt{\Delta p}$ but the fluctuation-dominated regime gives $q \sim \Delta p$. When interpreting measurements using a mean-pressure model, this change in scaling could be confused with viscous effects. 

$C_d$ values that were derived for models without fluctuation corrections should thus be interpreted with care, and they should not be used directly in a fluctuation-corrected model. Accounting for fluctuations in future model calibration will help distinguish these different mechanisms and improve the physical interpretation of calibrated airflow network parameters.

\section{Conclusion}
\label{sec::Conclusion}

This study demonstrates that geometry-specific surface pressures and fluctuation corrections can substantially improve airflow network predictions of natural ventilation. Using surface pressures from exterior-only LESs instead of standard ASHRAE pressure coefficients reduces the RMSE of predicted ventilation rates by approximately 50\%. This improvement largely arises because the LESs capture canopy flow effects that are poorly represented by ASHRAE shelter factors. The fluctuation corrections provide a further 10--40\% reduction in RMSE for most opening types and ventilation configurations by accounting for the nonlinear effects of pressure fluctuations on the mean ventilation rates.

The results also demonstrate that relatively simple implementations of the fluctuation corrections can be effective. Corrections based on pressure fluctuations are particularly attractive because they require only exterior flow information. While opening-specific $\sigma_{\Delta C_p}$ corrections implemented directly in the airflow network generally perform best, using bulk fluctuation intensities and applying the corrections after solving the network provides comparable accuracy in most cases, offering a practical alternative for existing airflow network models. For the building and urban canopy geometries considered in this study, we find that a discharge coefficient of $C_d = 0.7$ in combination with the fluctuation-corrected models approximately minimizes the RMSE. Bayesian calibration results further suggest that uncertainty in $C_d$ is not a dominant source of the remaining prediction errors.

In summary, this work demonstrates the potential of airflow networks to accurately predict natural ventilation rates across different interior layouts by accounting for urban canopy effects and pressure fluctuations. This pressure information comes from exterior-only LESs which do not resolve the indoor flow. Developing more efficient methods for estimating geometry-specific surface pressures and representative pressure fluctuation intensities will be important for implementing these improvements at scale. Future work should also investigate the transferability of the calibrated $C_d$ values for fluctuation-corrected models to other opening and building geometries. Progress in these directions may enable broadly accurate predictions for natural ventilation and its potential to offset mechanical cooling without relying on expensive simulations tools like LES.

\section*{CRediT authorship contribution statement}

\noindent\textbf{Nicholas Bachand:} Conceptualization, Methodology, Software,
Validation, Formal analysis, Investigation, Data curation, Visualization,
Writing -- original draft.
\textbf{Hesam Salehipour:} Supervision, Writing -- review and editing.
\textbf{Catherine Gorl\'{e}:} Conceptualization, Supervision, Resources,
Funding acquisition, Writing -- review and editing.

\section*{Declaration of competing interest}

The authors declare the following potential competing interests. H. Salehipour is
employed by Autodesk Research, which supported this work through a gift to
Stanford University. N. Bachand was previously an intern at Autodesk Research, working on projects unrelated to this study. The authors have no other known competing financial
interests or personal relationships that could have appeared to influence the
work reported in this paper.

\section*{Funding}

Funding: this work was supported by a gift from Autodesk Research and by
Stanford University's Strategic Energy Research Consortium. The funders had no role in the study design, analysis, interpretation of results, writing of the manuscript or the decision to submit, beyond H. Salehipour’s contributions as a co-author.

\section*{Data availability}

Pressure and ventilation rate data from the LESs are available in the Stanford Digital Repository (SDR). Data from the original coupled interior-exterior LESs of \citet{bachand_simulating_2026} are available at \url{https://doi.org/10.25740/vb900bf9392}, and data from the additional skylight and exterior-only LESs of this work are available at \url{https://doi.org/10.25740/bw345vw9929}. The airflow network code is available as the Python package \texttt{pyAFN} (version 0.1.0) in the SDR at \url{https://doi.org/10.25740/gz489mt0956} and on GitHub at \url{https://github.com/nbachand/pyAFN}. Additional data and analysis scripts are available from the corresponding author on request.

\section*{Declaration of generative AI and AI-assisted technologies in the manuscript preparation process} While preparing this work the authors used Anthropic’s Claude to reorganize and draft small portions of the text, check the manuscript for errors and internal consistency, separate appendix material from the main text, and apply the \LaTeX{} template. After using this tool, the authors reviewed and edited the content as needed and take full responsibility for the content of the published article.

\section*{Acknowledgements}

Computing for this project was performed on the Sherlock cluster. We thank
Stanford University and the Stanford Research Computing Center for providing
computational resources and support that contributed to these research results.

\clearpage

\bibliography{referencesZotero}

\clearpage

\appendix

\setcounter{equation}{0}\renewcommand{\theequation}{A\arabic{equation}}

\section{Derivations of the Fluctuation Corrections}
\label{app:supplementary}

\subsection{Mean-Dominated Regime}
\label{app::MeanDominated}
To derive the relationship between the time-averaged quantities in the mean-dominated regime we first square both sides of~\eqref{eq:inst_flow}:
\begin{equation}
(\overline{q} + q')^2 = \left( S C_{d} A \sqrt{\frac{2 |\overline{\Delta p} + \Delta p'|}{\rho}} \right)^2
\end{equation}
Expanding the squared terms gives:
\begin{equation}
(\overline{q})^2 + 2\overline{q}q' + (q')^2 = S^2 (C_d A)^2 \frac{2 |\overline{\Delta p} + \Delta p'|}{\rho}
\label{app_eq:squared_expanded}
\end{equation}
Next, we take the time-average, denoted by $\overline{(\cdot)}$, of each term in~\eqref{app_eq:squared_expanded}, and we remove the term $S^2 = (\operatorname{sgn}(\overline{\Delta p} + \Delta p'))^2$, which simplifies to 1 for any non-zero pressure drop:
\begin{equation}
\overline{ (\overline{q})^2 + 2\overline{q}q' + (q')^2 } = \overline{ \left( (C_d A)^2 \frac{2 |\overline{\Delta p} + \Delta p'|}{\rho} \right) }
\label{app_eq:rd_expanded}
\end{equation}
After removing terms that time-average to zero, the left-hand side simplifies to:
\begin{equation}
\overline{(\overline{q})^2} + 2\overline{q}\overline{q'} + \overline{(q')^2} = (\overline{q})^2 + \overline{(q')^2}
\label{app_eq:avg_lhs}
\end{equation}
In the mean-dominated regime, we assume the pressure difference never changes sign, i.e.\ $\operatorname{sgn}(\overline{\Delta p} + \Delta p') = \operatorname{sgn}(\overline{\Delta p})$, so the absolute value can be removed. The average of the pressure term then becomes:
\begin{equation}
\overline{|\overline{\Delta p} + \Delta p'|} = \overline{|\overline{\Delta p}| + \operatorname{sgn}(\overline{\Delta p})\,\Delta p'} = |\overline{\Delta p}| + \operatorname{sgn}(\overline{\Delta p})\,\overline{\Delta p'} = |\overline{\Delta p}|
\label{app_eq:avg_p}
\end{equation}
Substituting the results of Reynolds averaging into~\eqref{app_eq:rd_expanded} yields a relationship between the time-averaged quantities:
\begin{equation}
(\overline{q})^2 + \overline{(q')^2} = (C_d A)^2 \frac{2 |\overline{\Delta p}|}{\rho} = (q_{\text{MP}})^2
\label{app_eq:final_avg_flow_sqrd}
\end{equation}
Rearranging for $\overline{q}$ gives the mean-dominated correction,~\eqref{eq:qAFN_qstd}, with the fluctuation intensity $I$ defined in~\eqref{eq:FQ_qstd}.

We can alternatively approximate $I$ as a function of pressure fluctuations. 
Assuming the pressure fluctuations are small relative to the mean ($|\Delta p'| \ll |\overline{\Delta p}|$), we can perform a second-order binomial expansion of the instantaneous flow equation,~\eqref{eq:inst_flow}:
\begin{equation}
\begin{split}
\overline{q} + q' &= S C_d A \sqrt{\frac{2|\overline{\Delta p}|}{\rho}} \left(1 + \frac{\Delta p'}{\overline{\Delta p}}\right)^{1/2} \\
&\approx q_{\text{MP}} \left(1 + \frac{1}{2}\frac{\Delta p'}{\overline{\Delta p}} - \frac{1}{8}\bigg(\frac{\Delta p'}{\overline{\Delta p}}\bigg)^2\right)
\end{split}
\label{app_eq:second_order_expansion}
\end{equation}
Time averaging leaves:

\begin{equation}
\overline{q} \approx q_{\text{MP}} \left(1  - \frac{1}{8}\overline{\bigg(\frac{\Delta p'}{\overline{\Delta p}}\bigg)^2}\right)
\label{app_eq:second_order_expansion_averaged}
\end{equation}
These remaining terms are the first-order binomial expansion of:

\begin{equation}
\overline{q} \approx q_{\text{MP}}\sqrt{1 - \frac{1}{4}\overline{\bigg(\frac{\Delta p'}{\overline{\Delta p}}\bigg)^2}}
\label{app_eq:fluctuation_relation}
\end{equation}
This expression is equivalent to~\eqref{eq:qAFN_qstd} but with $I$ approximated as a function of pressure fluctuations,~\eqref{eq:FQ_pstd}.

\subsection{Fluctuation-Dominated Regime}
\label{app::FluctDominated}

If fluctuations are much larger than the mean we can no longer assume that the sign of pressure does not reverse. In this fluctuation-dominated regime, we start by factoring out the pressure fluctuation term in~\eqref{eq:inst_flow}:
\begin{equation}
\overline{q} + q' = S C_d A \sqrt{\frac{2|\overline{\Delta p} + \Delta p'|}{\rho}} = SC_d A \sqrt{\frac{2|\Delta p'|}{\rho}} \sqrt{\left|1 + \frac{\overline{\Delta p}}{\Delta p'}\right|}
\end{equation}
For small $|\overline{\Delta p}/\Delta p'|$, using $S \approx \operatorname{sgn}(\Delta p')$ and a first-order binomial expansion gives:
\begin{equation}
\overline{q} + q' \approx \operatorname{sgn}(\Delta p') C_d A \sqrt{\frac{2|\Delta p'|}{\rho}} \left(1 + \frac{1}{2}\frac{\overline{\Delta p}}{\Delta p'}\right)
\label{app_eq:q_lowbound_expansion}
\end{equation}
Expanding and time averaging this expression, it becomes:
\begin{equation}
\overline{q} \approx C_d A \sqrt{\frac{2}{\rho}} \left(\overline{\operatorname{sgn}(\Delta p')\sqrt{|\Delta p'|}} + \overline{\operatorname{sgn}(\Delta p')\frac{\overline{\Delta p}\sqrt{|\Delta p'|}}{2\Delta p'}}\right)
\label{app_eq:q_lowbound_mean}
\end{equation}
The first term cancels to zero if we assume the zero-centered distribution of $\Delta p'$ is also symmetric. The second term simplifies as 
\begin{equation}
\overline{\operatorname{sgn}(\Delta p')\frac{\overline{\Delta p}\sqrt{|\Delta p'|}}{2\Delta p'}} = \overline{\frac{\overline{\Delta p} \cdot \operatorname{sgn}(\Delta p')\sqrt{|\Delta p'|}}{2 \cdot \operatorname{sgn}(\Delta p')\sqrt{|\Delta p'|}\sqrt{|\Delta p'|}}} = \overline{\frac{\overline{\Delta p} }{2 \sqrt{|\Delta p'|}}}
\end{equation}
 Plugging the second term back into the full equation for $\overline{q}$ gives: 
\begin{equation}
\overline{q} \approx 
C_d A \overline{\Delta p} \cdot \overline{\sqrt{\frac{1}{2\rho|\Delta p'|}}}
\end{equation}
which is the fluctuation-dominated correction,~\eqref{eq:q_linear_deltap}, with $H$ in the pressure form,~\eqref{eq:RH_p}.

If instead the flow fluctuations $q'$ are symmetric, we expand in $q$. Inverting~\eqref{eq:inst_flow} gives $\Delta p = \rho\, q|q|/(2C_d^2A^2)$, and time-averaging with~\eqref{eq:qAFN} gives
\begin{equation}
q_{\text{MP}}|q_{\text{MP}}| = \overline{q|q|}.
\label{app_eq:qq_identity}
\end{equation}
For small $|\overline{q}/q'|$, a first-order expansion gives $(\overline{q}+q')|\overline{q}+q'| \approx q'|q'| + 2\overline{q}|q'|$. The first term time-averages to zero for symmetric $q'$, leaving
\begin{equation}
\overline{q} \approx \frac{q_{\text{MP}}|q_{\text{MP}}|}{2\,\overline{|q'|}},
\end{equation}
which is~\eqref{eq:q_linear_deltap} with $H$ in the ventilation form~\eqref{eq:RH_q}.

The two forms of $H$ are normalized power means. Writing the power mean of order $r$ of a non-negative quantity $x$ as $M_r(x) = \big(\overline{x^{r}}\big)^{1/r}$,
\begin{equation}
H(q) = \frac{M_{1}(|q'|)}{|q_{\text{MP}}|}, \qquad
H(\Delta p) = \frac{M_{-1}\big(\sqrt{|\Delta p'|}\big)}{\sqrt{|\overline{\Delta p}|}}
\label{app_eq:H_power_means}
\end{equation}
Because power means $M_r$ increase with order $r$, $M_{1}(|q'|)\le M_{2}(|q'|)=\sigma_q$, giving~\eqref{eq:RH_bound_q}, and $M_{-1}(\sqrt{|\Delta p'|})\le M_{4}(\sqrt{|\Delta p'|})=\sigma_{\Delta p}^{1/2}$, giving~\eqref{eq:RH_bound_p}.

\subsection{Closed-Form \texorpdfstring{$H$}{H} for Gaussian Fluctuations}
\label{app::GaussianH}

The Monte Carlo cases of Section~\ref{sssec::M-MC-Setup} use Gaussian fluctuations, for which $H$ and its models admit closed forms. These give an exact check on the numerics and quantify how tightly $H_{\mathrm{T}}$ tracks $H$ for both formulations.

For $X \sim \mathcal{N}(0,\sigma^{2})$ the absolute moments are
\begin{equation}
\overline{|X|^{s}} = \frac{\left(2\sigma^{2}\right)^{s/2}}{\sqrt{\pi}}\,
\Gamma\!\left(\frac{s+1}{2}\right), \qquad s > -1
\label{app_eq:absmoment}
\end{equation}
The pressure form of $H$,~\eqref{eq:RH_p}, requires $s=-1/2$:
\begin{equation}
\overline{|\Delta p'|^{-1/2}}
= \kappa\,\sigma_{\Delta p}^{-1/2}, \qquad
\kappa = \frac{\Gamma(1/4)}{2^{1/4}\sqrt{\pi}} \approx 1.7201
\label{app_eq:kappa}
\end{equation}
Substituting into~\eqref{eq:RH_p} and using $2I(\Delta p)=\sigma_{\Delta p}/|\overline{\Delta p}|$,
\begin{equation}
H(\Delta p) = \frac{1}{\kappa}\sqrt{\frac{\sigma_{\Delta p}}{|\overline{\Delta p}|}}
= \frac{\sqrt{2}}{\kappa}\sqrt{I(\Delta p)} \approx 0.8222\,\sqrt{I(\Delta p)}
\label{app_eq:H_gauss}
\end{equation}
The pressure-form models, $H_{\mathrm{U}}(\Delta p)=\sqrt{2I(\Delta p)}$ from~\eqref{eq:RH_bound_p} and $H_{\mathrm{T}}(\Delta p)=\sqrt{(3\sqrt{3}/8)I(\Delta p)}$ from~\eqref{eq:HT_p}, share the same $\sqrt{I}$ scaling as $H(\Delta p)$, so each model's ratio to the exact $H$ is a constant independent of $I$, $\sigma_{\Delta p}$ and $\overline{\Delta p}$.

For Gaussian flow fluctuations,~\eqref{app_eq:absmoment} with $s=1$ gives $\overline{|q'|} = \sqrt{2/\pi}\,\sigma_q$, so the ventilation form is $H(q) = \sqrt{2/\pi}\,I(q)$ exactly. Its models are again constant multiples of $H$.

\subsubsection{Upper Bound}

For the pressure form, the ratio of the upper bound~\eqref{eq:RH_bound_p} to the exact result~\eqref{app_eq:H_gauss} is
\begin{equation}
\frac{H_{\mathrm{U}}}{H} = \kappa = \frac{\Gamma(1/4)}{2^{1/4}\sqrt{\pi}} \approx 1.7201
\label{app_eq:HU_ratio}
\end{equation}
so the upper bound is never tight, exceeding the exact $H$ by $72.0\%$ for every finite $I>0$. For the ventilation form, $H_{\mathrm{U}}(q)/H(q) = \sqrt{\pi/2} \approx 1.2533$, so the bound exceeds the exact $H$ by $25.3\%$.

\subsubsection{Tangent-Intersect Model}

In the pressure form, the corresponding ratio for the tangent-intersect model is
\begin{equation}
\frac{H_{\mathrm{T}}}{H} = \frac{3^{3/4}\,\Gamma(1/4)}{2^{9/4}\sqrt{\pi}} \approx 0.9802
\label{app_eq:HT_ratio}
\end{equation}
The tangent-intersect model therefore underpredicts the exact $H$ by $1.98\%$ for every finite $I>0$. Since $\overline{q} \propto 1/H$, this corresponds to a $2.02\%$ overprediction of $\overline{q}$ wherever the fluctuation-dominated form~\eqref{eq:q_linear_deltap} applies. For the ventilation form, $H_{\mathrm{T}}(q) = H_{\mathrm{U}}(q)$ uses the root-mean-square of $q'$ rather than its mean magnitude, so the tangent-intersect model overpredicts the exact $H$ by $25.3\%$, a $20.2\%$ underprediction of $\overline{q}$. These values are the large-$I$ limit ($I^{-1} \to 0$). For the pressure form, the piecewise model's error is larger near the transition, peaking at about 9\% just above $(I^*)^{-1}$, where the mean-dominated branch applies; for the ventilation form, the error decreases monotonically from $20.2\%$ toward zero through the transition.

\end{document}